\documentclass[%
	reprint,
	superscriptaddress,
	amsmath,amssymb,amsfonts,
	aps,
	pra,
]{revtex4-2}

\usepackage[utf8]{inputenc}
\usepackage[T1]{fontenc}
\usepackage{lmodern}
\usepackage[english]{babel}
\usepackage[autostyle=true]{csquotes}

\DeclareUnicodeCharacter{030C}{\v{}}
\DeclareUnicodeCharacter{0301}{\'{}}

\usepackage[dvipsnames]{xcolor}
\usepackage{graphicx}
\graphicspath{{figures/}}

\usepackage{tikz}
\usepackage{tikz-cd}
\usepackage[normalem]{ulem}
\usepackage{mhchem}
\usetikzlibrary{%
  matrix,%
  calc,%
  arrows%
}
\usetikzlibrary{quantikz2} 

\usepackage{amssymb,amsmath}

\usepackage{braket}
\usepackage{bm}
\usepackage{mathtools}
\usepackage{braket}

\usepackage{xspace}
\usepackage{extdash}
\usepackage{mleftright}
\mleftright

\usepackage{hyperref}

\usepackage[
	capitalise,
]{cleveref}

\usepackage{acro}
\acsetup{%
	first-style=long-short,
}

\usepackage{url}

\definecolor{mypurple}{rgb}{0.49,0.18,0.56}
\definecolor{mygold}{rgb}{0.93,0.49,0.13}
\definecolor{mygreen}{rgb}{0,0.5,0}
\definecolor{myblue}{rgb}{0,0,0.75}
\definecolor{mymagenta}{cmyk}{0,1,0,0.12}
\definecolor{mygray}{rgb}{0.5,0.5,0.5}

\newif\ifcomments
\commentstrue 

\begin{document}

\title{Quantum resources in non-stoquastic quantum annealing}

\author{Chiara Capecci}
\thanks{These authors contributed equally to this work.}
\affiliation{Pitaevskii BEC Center and Department of Physics, University of Trento, Via Sommarive 14,
38123 Trento, Italy}
\affiliation{INFN-TIFPA, Trento Institute for Fundamental Physics and Applications, Trento, Italy}

\author{Sebastian Nagies}
\thanks{These authors contributed equally to this work.}

\affiliation{Pitaevskii BEC Center and Department of Physics, University of Trento, Via Sommarive 14,
38123 Trento, Italy}
\affiliation{INFN-TIFPA, Trento Institute for Fundamental Physics and Applications, Trento, Italy}

\author{Naga Dileep Varikuti}
\affiliation{Pitaevskii BEC Center and Department of Physics, University of Trento, Via Sommarive 14,
38123 Trento, Italy}
\affiliation{INFN-TIFPA, Trento Institute for Fundamental Physics and Applications, Trento, Italy}

\author{Philipp Hauke}
\affiliation{Pitaevskii BEC Center and Department of Physics, University of Trento, Via Sommarive 14,
38123 Trento, Italy}
\affiliation{INFN-TIFPA, Trento Institute for Fundamental Physics and Applications, Trento, Italy}

\begin{abstract}

Quantum annealing promises to solve combinatorial optimization problems by preparing the ground state of a target Hamiltonian. Standard annealing protocols are, however, stoquastic and can thus be simulated by sign-problem-free quantum Monte-Carlo methods. To obtain a true quantum advantage, it has been proposed to use non-stoquastic catalyst Hamiltonians. Active only at intermediate stages of the protocol, these can, for certain problems, convert first-order into second-order quantum phase transitions and thus permit an exponential speedup over the stoquastic protocol. At the same time, the non-stoquastic catalyst renders quantum Monte-Carlo methods inefficient. It remains, however, an open question how other classical computation methods are affected by the non-stoquastic terms. We address this question by computing quantum resources---entanglement entropy and stabilizer Rényi entropy---whose presence makes classical computations based on tensor networks and stabilizer-tableau methods exponentially hard. We compare these with the spectral gap along the annealing path for two paradigmatic benchmark models, the fully connected $p$-spin model and a geometrically local Ising model. 
While the exact behavior shows a subtle dependency on the underlying model and the annealing path, our numerics suggest consistently that the scaling of entanglement and non-stabilizerness is at least maintained in the deeply non-stoquastic regime and in some cases even significantly enhanced. 
Our results thus suggest that improvements of quantum performance in non-stoquastic annealing coincide with significant presence of quantum computational resources.

\end{abstract}

\maketitle

\section{Introduction}\label{sec:intro}

Quantum annealing is a promising paradigm for solving classical combinatorial optimization problems and preparing ground states of many-body systems, in particular spin models~\cite{Santoro2002, Martonak2004, Albash2018, Hauke2020, Rajak2022, Nagies2025a, Nagies2026c}. The basic idea is to initialize the system in an easily prepared ground state of a driver Hamiltonian, and then slowly interpolate toward the problem Hamiltonian whose ground state encodes the solution. If the interpolation is sufficiently slow, the adiabatic theorem guarantees that the state tracks the instantaneous ground state along the evolution, so that the target state is obtained at the end of the protocol. The annealing time required for adiabaticity scales polynomially with the inverse minimum energy gap between the ground state and the first excited state~\cite{Jansen2007,Lidar2009,Amin2009,Cheung2011}.
In standard implementations of quantum annealing, the problem is encoded in a classical Ising model as the cost Hamiltonian and employs a transverse-field driver, making the total annealing Hamiltonian stoquastic~\cite{Bravyi2006}.
Stoquastic Hamiltonians are free of the sign problem, which therefore poses no obstruction to their simulation via quantum Monte Carlo methods~\cite{Troyer2005, Isakov2016,  Denchev2016, Jiang2017, Hormozi2017}.
Evidence from the study of complexity classes suggests that non-stoquasticity may be required for achieving a genuine quantum advantage with adiabatic quantum computation~\cite{kitaev2002classical, Kempe2004, Bravyi2006, Bravyi2010}. This has motivated growing interest in non-stoquastic annealing protocols, where catalyst Hamiltonians are introduced that vanish at the start- and endpoints of the annealing evolution~\cite{Farhi2000, Nishimori2017, Hormozi2017, Albash2018, Albash2019} such that they break stoquasticity without modifying initial and final state.

Non-stoquastic terms in the Hamiltonian can substantially modify the instantaneous energy spectrum of the annealing protocol and, in particular, the nature of the quantum phase transitions encountered along the evolution~\cite{Seki2012, Nishimori2017, Albash2019}. For the paradigmatic $p$-spin model, Seki and Nishimori~\cite{Seki2012} showed that antiferromagnetic transverse catalysts can turn the first-order transition of the stoquastic protocol, with its exponentially closing minimum energy gap, into a second-order transition with polynomial gap scaling, yielding an exponential speedup of the adiabatic evolution. 
This result suggests that non-stoquastic terms may be a key ingredient for achieving a potential quantum speedup through quantum optimization \cite{Seki2012, Hormozi2017, Nishimori2017, Albash2019, Hegade2022}. 
However, a genuine quantum advantage requires that the state cannot be efficiently tracked by any classical method, including besides quantum Monte Carlo \cite{Suzuki1993,Troyer2005, Becca2017} also, e.g., tensor networks \cite{Schollwoeck2011, Orus2014, Montangero} and stabilizer-tableau simulation \cite{gottesman1998heisenberg, aaronson2004improved}. The difficulty of representing a given quantum state through the latter two methods is quantified by its quantum resource content in terms of entanglement and non-stabilizerness (also called magic), respectively. For stoquastic sweeps, it is by now well established that quantum annealing has to pass through an `entanglement barrier' \cite{Hauke2015, dupont2022entanglement, chen2022much, Santra2025}, and recently the existence of a similar `magic barrier' has been identified \cite{Capecci2025}. 
However, it remains an important open question how the introduction of non-stoquastic terms impacts these quantum resources. In particular, one may wonder whether an enlarging of the spectral gap would be accompanied by a reduction of quantum resources, as this would imply that classical simulation methods become easier and thus counter-act any potential quantum speedup. 

To address this question, we compare stoquastic and non-stoquastic annealing protocols by monitoring, along the evolution,  on the one hand the minimum energy gap that controls the adiabatic runtime and, on the other hand, two complementary quantum resources of the instantaneous ground state: entanglement~\cite{Amico2008entanglement, Horodecki2009entanglement, contreras2019resource, bauml2019resource} and non-stabilizerness~\cite{gottesman1997stabilizer,gottesman1998heisenberg, Briegel2009, veitch2014resource, campbell2017, leone2024stabilizer}. 
Together, they provide complementary perspectives on the role of non-stoquasticity, highlighting two distinct ways in which a quantum state can become difficult to simulate classically.

We carry out this analysis on two complementary benchmark models. The first is the above-mentioned fully connected ferromagnetic $p$-spin model~\cite{Seki2012}. The permutation symmetry of this model allows both an analytical treatment in the thermodynamic limit and numerical simulation at large system sizes, which makes the model---even if somewhat artificial---an important benchmark problem \cite{Joerg2010,Bapst2012,Wauters2017, Wauters2020}. 
The second is a specific geometrically local Ising model introduced by Albash~\cite{Albash2019}, for which numerical evidence hints at a similar exponential advantage of non-stoquastic quantum annealing. Its local connectivity makes the model more representative of near-term hardware implementations, at the cost of restricting the analysis to smaller sizes accessible by exact diagonalization. 
In both models, we find that entanglement and non-stabilizerness show prominent features at the gap closing. 
Moreover, in none of the considered scenarios does the increase of the gap due to the non-stoquastic catalyst reduce the considered quantum resources, though the exact behavior depends on the model and considered annealing path. 
Specifically, entanglement saturates in the stoquastic regimes of the studied models, but in non-stoquastic regimes it can reach significantly higher values and continue to grow over all studied system sizes. 
In contrast, the non-stabilizer Rényi entropy is extensive in all considered scenarios, but in non-stoquastic regimes it can achieve significantly larger scaling than in stoquastic regimes. 
These findings indicate that classical computation based on tensor networks and stabilizer-tableau methods can face increased difficulties when simulating non-stoquastic annealing. 
Our analysis thus mirrors, at the level of quantum resources, the complexity-theoretic arguments that motivate non-stoquasticity.

This paper is organized as follows: In Sec.~\ref{sec:background}, we introduce the theoretical background of quantum annealing, non-stoquastic Hamiltonians and the different quantum resources considered in this work. In Sec.~\ref{sec:pspin}, we study the $p$-spin model, investigating the interplay between spectral properties and quantum resources. In Sec.~\ref{sec:localising}, we extend the analysis to the geometrically local Ising model, providing a more realistic setting for non-stoquastic annealing. 
Finally, in Sec.~\ref{sec:conclusion}, we summarize the main findings and discuss their implications for non-stoquastic quantum annealing.

\section{Background}\label{sec:background}
We organize the background in two parts. Sec.~\ref{subsec:annealing_background} reviews quantum annealing and the complexity-theoretic motivation for non-stoquastic catalyst Hamiltonians, while Sec.~\ref{subsec:resources_background} introduces the two quantum resources considered in this work, bipartite entanglement entropy and stabilizer Rényi entropy, whose interplay with the spectral gap is the central object of study.

\subsection{Quantum annealing with non-stoquastic Hamiltonians} \label{subsec:annealing_background}

This section reviews the standard quantum annealing protocol and its performance limitations imposed by the minimum spectral gap. It introduces the notion of stoquastic Hamiltonians, the complexity-theoretic constraints they impose on adiabatic quantum computation, and discusses catalyst Hamiltonians as a route to non-stoquastic annealing schedules that may overcome these constraints.

\vspace{0.3cm}
\textit{Standard quantum annealing protocol} --- Quantum annealing is a family of algorithms for solving combinatorial optimization problems that exploits quantum fluctuations to search for the ground state of a problem Hamiltonian~\cite{Farhi2000, Albash2018,Hauke2020}. The optimization problem is encoded in a cost Hamiltonian $H_\mathrm{cost}$, chosen to be diagonal in the computational basis such that its ground state corresponds to the optimal solution. To reach this target state, the system is first prepared in the ground state of a simple driving Hamiltonian $H_\mathrm{drive}$, typically a uniform transverse field $H_\mathrm{drive} = -\sum_i \hat{\sigma}_i^x$, whose ground state $\ket{+}^{\otimes N}$ is easy to prepare. The system then evolves under the time-dependent Hamiltonian
\begin{equation}\label{eq:standard_annealing}
    H(s) = (1-s)H_\mathrm{drive} + sH_\mathrm{cost},
\end{equation}
where $s = t/t_\mathrm{sweep} \in [0,1]$ parametrizes the normalized time along the annealing path. At $s = 0$, the Hamiltonian coincides with $H_\mathrm{drive}$, and at $s = 1$ it reduces to $H_\mathrm{cost}$.
The adiabatic theorem~\cite{Jansen2007, Lidar2009, Amin2009} guarantees that the system remains in the instantaneous ground state of $H(s)$ throughout the evolution, provided the sweep is performed sufficiently slowly. Specifically, the required sweep time scales as $t_\mathrm{sweep} \propto \Delta E^{-2}$, where $\Delta E$ is the minimum energy gap between the instantaneous ground state and first excited state encountered along the annealing path~\cite{Hauke2020, Rajak2022}. For hard optimization problems, this gap typically closes exponentially with the system size, $\Delta E \propto e^{-\alpha N}$,  as commonly observed near first-order quantum phase transitions or in spin-glass phases, reducing the practicality of the protocol due to exponentially long annealing times~\cite{Joerg2010, Heim2015}.

\vspace{0.3cm}
\textit{Stoquastic Hamiltonians} --- 
The name \emph{stoquastic} was introduced in Ref.~\cite{Bravyi2006}, where the authors define a stoquastic Hamiltonian as one whose off-diagonal matrix elements are real and non-positive in the computational basis. By this definition, the standard quantum annealing Hamiltonian in Eq.~\eqref{eq:standard_annealing} is stoquastic. Stoquastic Hamiltonians do not suffer from the sign problem in quantum Monte Carlo (QMC) methods~\cite{Troyer2005}, as their partition function decomposition involves only non-negative weights. However, whether a given non-stoquastic Hamiltonian can be made stoquastic by local basis changes is itself computationally hard in general~\cite{Marvian2019, Klassen2020}, complicating the use of stoquasticity as a practical diagnostic for QMC simulability. Moreover, the absence of a sign problem does not always guarantee efficient classical simulation, since finding the ground state of even classical Ising models is \textbf{NP}-hard in general~\cite{Barahona1982}.

In the framework of Hamiltonian complexity, the local Hamiltonian problem (LH-MIN) ~\cite{kitaev2002classical,Kempe2004} for stoquastic Hamiltonians is contained in the complexity class \textbf{AM}, a probabilistic generalization of \textbf{NP}. Moreover, it is \textbf{MA}-hard for the stoquastic 2-local case~\cite{Bravyi2006}, and is \textbf{MA}-complete in the frustration-free case~\cite{Bravyi2010}. For stoquastic Hamiltonians with a polynomial spectral gap, LH-MIN is further contained in \textbf{PostBPP}, implying that efficient adiabatic quantum computation restricted to stoquastic Hamiltonians is contained within this classical complexity class, which lies at a low level of the polynomial hierarchy~\cite{Bravyi2006}. This suggests that non-stoquastic Hamiltonians may be necessary for universal quantum computational advantage 
(although a superpolynomial speedup of quantum versus classical algorithms has been demonstrated even for cases of sign-problem-free adiabatic computation~\cite{Gilyen2021}). 
Concretely, the authors of Ref.~\cite{Biamonte2008a} identify minimal two-local qubit Hamiltonians that restore \textbf{QMA}-completeness: the transverse-field Ising model supplemented by transversal two-body $\hat\sigma_i^x \hat\sigma_j^x$ interactions, as well as a second model consisting of longitudinal and transverse fields with two-body $\hat\sigma_i^z \hat\sigma_j^x$ couplings.

\begin{figure}\label{fig:schedule_sketch}
    \centering
    \includegraphics[width=0.75\linewidth]{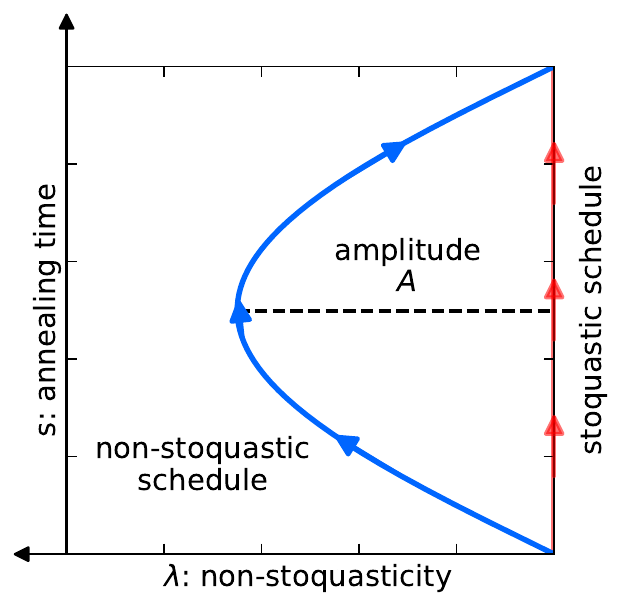}
    \caption{Illustration of two types of annealing schedules in the $(\lambda, s)$ parameter space. The red line ($\lambda (s) = 0$) represents a conventional stoquastic schedule, where only the normalized annealing time $s$ varies. The blue curve depicts a non-stoquastic schedule, in which the strength $\lambda$ of the non-stoquastic terms in the annealing Hamiltonian are tuned by the amplitude $A$. Schedules of this type are employed for the $p$-spin model in Sec.~\ref{sec:pspin}, cf.\ Eq.~\eqref{eq:pspin_schedule}, where they can permit to circumvent a line of first-order phase transitions~\cite{Nishimori2017}.}
    \label{fig:schedule_sketch}
\end{figure}

\vspace{0.3cm}
\textit{Catalyst Hamiltonians} ---
A natural strategy to move beyond the computational limitations that stoquasticity imposes on the standard annealing protocol is to modify the annealing path by introducing an additional \emph{catalyst} Hamiltonian $H_\mathrm{catalyst}$~\cite{Farhi2000,Albash2018}. The catalyst is chosen to be non-stoquastic in the computational basis, so that the total annealing Hamiltonian, which includes the stoquastic drive and cost terms, inherits this property and is itself non-stoquastic. To preserve the initial and final states of the protocol, the catalyst is taken to be active only during the intermediate stages of the evolution. The modified annealing Hamiltonian reads

\begin{equation}\label{eq:H_catalyst}
\tilde{H}(s) =  (1-s) H_\mathrm{drive} + \lambda ( s) H_\mathrm{catalyst} + s H_\mathrm{cost},
\end{equation}
where the schedule function $\lambda(s)$ controls the strength of the catalyst term. This function is chosen to satisfy $\lambda(0) = \lambda(1) = 0$, ensuring that the non-stoquastic contribution is active only during the intermediate stages of the anneal: the system is still initialized in the easily prepared ground state of $H_\mathrm{drive}$ at $s = 0$ and evolves toward the solution encoded in $H_\mathrm{cost}$ at $s = 1$. Both stoquastic and non-stoquastic annealing schedules are illustrated in Fig.~\ref{fig:schedule_sketch}.

The physical intuition behind the catalyst approach is that higher-order quantum fluctuations, introduced through terms such as two-body $\hat{\sigma}^x_i \hat{\sigma}^x_j$ or three-body $\hat{\sigma}^x_i \hat{\sigma}^x_j \hat{\sigma}^x_k$ interactions in $H_\mathrm{catalyst}$, can modify the energy landscape traversed during the anneal. In particular, these terms can couple the instantaneous ground and first excited states, potentially enlarging the minimum energy gap and thus reducing the sweep time required by the adiabatic theorem~\cite{Seki2012, Seki2014}. Alternatively, even when the minimum gap is not significantly increased, the catalyst may open secondary avoided crossings that enable \emph{diabatic} protocols, in which controlled transitions through narrow gaps are exploited rather than avoided~\cite{Feinstein2025, Ghosh2026}. 

However, the effectiveness of catalyst Hamiltonians is highly problem-dependent. Numerical studies on random Ising spin glasses~\cite{Hormozi2017} found that non-stoquastic drivers outperform their stoquastic counterparts only for a small fraction of harder instances. The authors in Ref.~\cite{Hormozi2017} present evidence suggesting that the observed improvement arises from promoting beneficial diabatic transitions rather than from a uniform increase of the spectral gap. Despite this mixed picture, there exist problems for which non-stoquastic catalysts provably yield exponential speedups. A canonical example is the ferromagnetic $p$-spin model, where a non-stoquastic catalyst converts a first-order quantum phase transition into a second-order one, replacing the exponential closing of the spectral gap with a polynomial one~\cite{Seki2012}. Further, Ref.~\cite{Albash2019} constructs explicit 2-local and geometrically local models that potentially exhibit exponential advantages for non-stoquastic catalysts. In this work, we study these models as important benchmarks for non-stoquastic quantum annealing.

\subsection{Quantum resources} \label{subsec:resources_background}

To understand the quantum behavior of the annealing protocols, we analyze two complementary quantum resources of the instantaneous ground state: entanglement and non-stabilizerness.
These two resources limit efficient classical simulation of quantum systems in complementary ways. States with area-law entanglement admit efficient tensor-network representations such as matrix product states~\cite{Schollwoeck2011, Orus2014}. On the contrary, stabilizer states, even when highly entangled, can be efficiently simulated classically 
due to the Gottesman--Knill theorem~\cite{gottesman1998heisenberg}. Only when both entanglement and non-stabilizerness are simultaneously large, both simulation strategies become exponentially costly. This motivates analyzing both quantities together to examine the difficulty of classical simulation during the annealing sweep.

\vspace{0.3cm}
\textit{Bipartite entanglement entropy}---For a pure state $\ket{\psi}$ of a bipartite system composed of subsystems $A$ and $B$, the bipartite entanglement entropy is obtained by first constructing the reduced density matrix $\rho_A = \mathrm{Tr}_B(\ket{\psi}\bra{\psi})$, where the partial trace is taken over subsystem $B$. The entanglement entropy is then given by the von Neumann entropy  of the reduced state \cite{Amico2008entanglement, Horodecki2009entanglement},
\begin{equation}
    S(\rho_A) = -\mathrm{Tr}(\rho_A \log_2 \rho_A) = -\sum_i \lambda_i \log_2 \lambda_i,
\label{eq:entanglement-entropy}
\end{equation}
where $\lambda_i$ are the eigenvalues of $\rho_A$. Equivalently, $S(\rho_A)$ can be computed directly from the Schmidt coefficients of $\ket{\psi}$ with respect to the bi-partition $A|B$. For a maximally entangled state of subsystems with dimensions $d_A$ and $d_B$, with $d=\min(d_A,d_B)$, the entanglement entropy attains its maximum value of $\log_2 d$.

\vspace{0.3cm}
\textit{Stabilizer Rényi entropy}---The stabilizer Rényi entropy (SRE) quantifies the non-stabilizerness \cite{BravyiKitaev2005} of a quantum state by measuring how it spreads over the basis of Pauli strings. For a pure  state $\ket{\psi}$ of $N$ qubits, the SRE is defined as~\cite{Leone2022,Bittel2026}
\begin{equation}
    M_{\alpha}(\ket{\psi}) = \frac{1}{1-\alpha}\log_2 \left[ \sum_{P \in \mathcal{P}_N} 
    \frac{\braket{\psi|P|\psi}^{2\alpha}}{2^N} \right],
\label{eq:SRE}
\end{equation}
where the sum runs over all $N$-qubit Pauli strings $\mathcal{P}_N = \{ \sigma_{i_1} \otimes \cdots \otimes \sigma_{i_N} \}$ with  $\sigma_i \in \{I, \sigma^x, \sigma^y, \sigma^z\}$. The SRE is non-negative and vanishes if and only if  $\ket{\psi}$ is a stabilizer state, and for Rényi index $\alpha \geq 2$, it is a monotone for magic-state resource theory restricted to pure states~\cite{leone2024stabilizer}. In this work, we always consider $\alpha = 2$, defining the Second stabilizer Rényi entropy $M_{2}$, which we refer to as SRE.

The stabilizer Rényi entropy has emerged in recent years as a key measure of non-stabilizerness that is relevant for quantum error correction~\cite{sierant2026, Spagnoli2026}, magic-state distillation~\cite{BravyiKitaev2005}, classical simulability~\cite{gottesman1998heisenberg, aaronson2004improved,Bravyi2016simulation}, and quantum chaos ~\cite{Leone2021, santra2025SYK, Magni2025quantumcomplexity, loio2025quantum, magni2025anticoncentration, varikuti2025deep,Varikuti2026}. The fact that SRE is a strict monotone makes it an interesting quantifier of non-stabilizerness~\cite{leone2024stabilizer}. It has been investigated in optimization algorithms where `magic barriers' have been observed~\cite{Capecci2025}.

\section{p-spin model}\label{sec:pspin}
In this section, we study the interplay between spectral properties and quantum resources in the non-stoquastic annealing protocol for the fully connected $p$-spin model~\cite{Seki2012}. We introduce the model Hamiltonian and its permutation-symmetric structure, enabling exact diagonalization in the Dicke subspace (Sec.~\ref{subsec:p_spin_definition+dicke}). We analyze the energy gap, bipartite entanglement, and stabilizer Rényi entropy across the annealing parameter space (Sec.~\ref{subsec:p-spin_parameterspace}) and their finite-size scaling along sinusoidal schedules (Sec.~\ref{subsec:p-spin_scaling}). Finally, we examine the behaviour of entanglement and magic along the annealing sweep (Sec.~\ref{subsec:p-spin_barriers}).

\subsection{Problem definition and Dicke basis}\label{subsec:p_spin_definition+dicke}

The ferromagnetic $p$-spin model ($p\geq 2$) is defined as
\begin{equation}
\label{eq:Hcost_pspin}
H_\mathrm{cost} = -N \bigg( \frac{1}{N} \sum_{i=1}^N \sigma_i^z \bigg)^p \,.
\end{equation}
The ground state of this Hamiltonian is the fully polarized state with all spins pointing up for odd $p$, while for even $p$ both fully polarized states are degenerate. For $p=2$, the model reduces to the Lipkin--Meshkov--Glick model~\cite{Lipkin1965, Meshkov1965, Glick1965}. Interestingly, in the limit $p\to\infty$ the energy landscape becomes flat for all states except the ground state manifold, effectively recovering the Grover search problem~\cite{Joerg2010}.
Despite being classically tractable, the $p$-spin model exhibits rich behavior, making it a valuable testbed for benchmarking quantum-optimization algorithms. Its fully connected interaction structure renders mean-field theory exact in the thermodynamic limit~\cite{Joerg2010, Bapst2012, Seki2012}, allowing for analytical treatment via semiclassical methods. The standard quantum annealing protocol with a transverse-field driver $H_\mathrm{drive} = -\sum_i \sigma_i^x$ encounters a first-order quantum phase transition for $p \geq 3$, where the minimum energy gap closes exponentially with system size ~$N$ ~\cite{Joerg2010}. 
Therefore, the required runtime of the protocol increases exponentially with $N$.

Seki and Nishimori ~\cite{Seki2012} showed that this first-order transition can be circumvented by introducing antiferromagnetic quantum fluctuations through the catalyst Hamiltonian
\begin{equation}
\label{eq:Hcat_pspin}
H_\mathrm{catalyst} = +N \bigg( \frac{1}{N} \sum_{i=1}^N \sigma_i^x \bigg)^2 \,,
\end{equation}
which constitutes non-stoquastic all-to-all two-body interactions along the transverse direction. The total time-dependent annealing Hamiltonian then takes the form
\begin{equation}
\label{eq:H_nonstoquastic_pspin}
\hat{H}(s, \lambda) = s\lambda\, H_\mathrm{cost} + s(1-\lambda)\, H_\mathrm{catalyst} + (1-s)\, H_\mathrm{drive} \,,
\end{equation}
where the parameters $s$ and $\lambda$ vary in time, starting from $(s, \lambda) = (0, \text{arbitrary})$ and ending at $(s, \lambda) = (1, 1)$. For intermediate values of $5 \leq p \leq 21$, a path through the parameter space exists that avoids all first-order phase transitions~\cite{Seki2012}. Along this path the exponentially closing gap gets replaced by a polynomially closing one, signaling a second-order transition point~\cite{Seki2012, Nishimori2017}. According to the adiabatic theorem, this leads to an exponential enhancement of the annealing performance. However, in the limit $p \to \infty$, first-order transitions become unavoidable, consistent with the optimality of the Grover bound \cite{Grover1997,Joerg2010}.

\textit{Numerical methods}--- 
The key feature enabling efficient numerical treatment of the $p$-spin model is its permutation symmetry: since all terms in the Hamiltonian depend only on the collective magnetization $\sum_i \sigma_i^z$, the Hamiltonian is invariant under arbitrary permutations of the spins. Consequently, the dynamics can be restricted to the fully symmetric subspace of the $N$-qubit Hilbert space, spanned by the $N+1$ Dicke states $\ket{S, m}$ with $S = N/2$.
In practice, we perform exact diagonalization in the Dicke basis to obtain instantaneous eigenstates and eigenvalues of the Hamiltonian in Eq.~\eqref{eq:H_nonstoquastic_pspin} for fixed values of $s$ and $\lambda$. From these we extract the energy gap and compute the bipartite entanglement entropy and the second stabilizer Rényi entropy of the instantaneous ground state. The entanglement is evaluated for a balanced bipartition with $n_A = n_B = N/2$, and can be obtained directly from the Dicke coefficients \cite{Popkov2012}. On the other hand, for the SRE, one can group the Pauli strings  into equivalence classes labeled by the number of $I$, $\sigma^x$, $\sigma^y $, and $\sigma^z$ operators they contain, reducing the computational cost to polynomial scaling in system size~\cite{Stockton2003, Passarelli2024} (see Appendix~\ref{app:p-spin} for details).
This approach allows us to access system sizes of more than $100$ qubits.

\begin{figure*}
    \centering
    \includegraphics[width=1\linewidth]{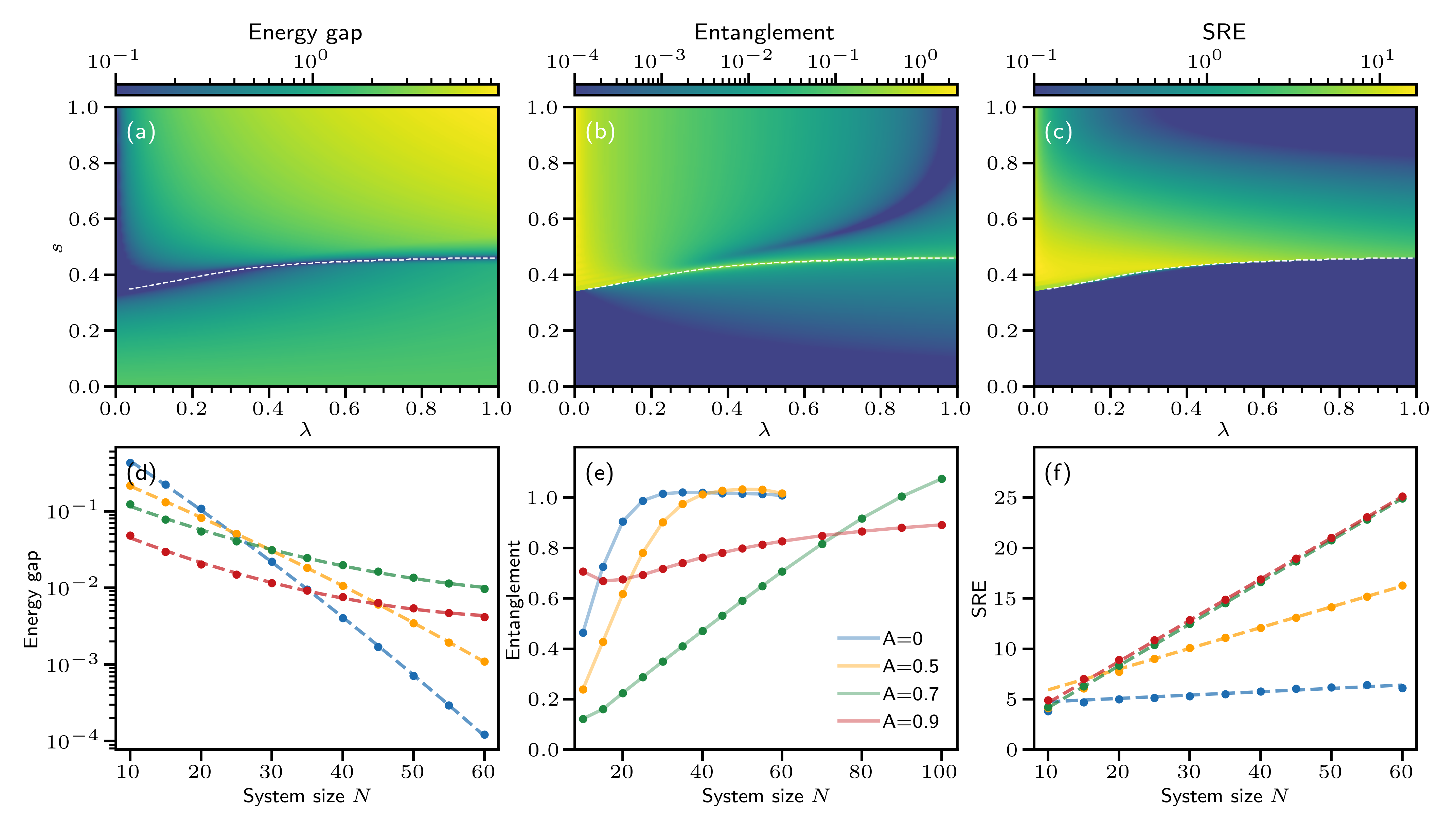}
    \caption{\textbf{First row}: (a) energy gap, (b) bipartite entanglement entropy, and (c) stabilizer Rényi entropy as functions of the annealing parameters $s$ and $\lambda$, for a $p$-spin model cost Hamiltonian with $N=30$ and $p=5$. All observables are evaluated on the instantaneous ground state of the Hamiltonian in Eq.~\eqref{eq:H_nonstoquastic_pspin}. All quantities are shown in logarithmic scale to enhance the visibility of relevant features across the annealing landscape. 
    Both entanglement entropy and stabilizer Rényi entropy show a pronounced peak where the minimum gap occurs. 
    \textbf{Second row}: finite-size scaling of (d) the minimum energy gap, (e) the maximum bipartite entanglement entropy, and (f) the maximum stabilizer Rényi entropy along annealing paths defined by $\lambda(s)=1-A\sin(\pi s)$, for amplitudes $A=0,\,0.5,\,0.7,\,0.9$. For each schedule, the minimum gap and the maximum values of entanglement and SRE are extracted along the sweep and plotted as functions of the system size $N$.
    Dashed lines in panel (d) correspond to fits of the minimum energy gap. For $A=0$ and $A=0.5$, we use exponential fits of the form $ \Delta E_{\min} \approx \exp( a + b N)$, with $(a,b)_{A=0}=(0.597,-0.134)$ and $(a,b)_{A=0.5}=(-0.666,-0.085)$. For $A=0.7$ and $A=0.9$, we employ exponential fits of the form $\Delta E_{\min} \approx \exp( a + b N + c N^2)$, with $(a,b,c)_{A=0.7}=(-1.34,-8.68\times10^{-2},\,5.43\times10^{-4})$ and $(a,b,c)_{A=0.9}=(-2.23,-9.46\times10^{-2},\,6.84\times10^{-4})$.
    Solid lines in panel (e) are guides to the eye.
    In panel (f), dashed lines show linear fits of the maximum SRE for $N \geq 30$ of the form $M_2^{\max} = a + b N$, with $(a,b)_{A=0}=(4.42,\,0.033)$, $(a,b)_{A=0.5}=(3.87,\,0.206)$, $(a,b)_{A=0.7}=(0.019,\,0.415)$, and $(a,b)_{A=0.9}=(0.563,\,0.409)$. 
    The scaling behavior indicates that the regime with improved annealing performance coincides with enhanced growth of non-stabilizerness.
    By contrast, for the considered system sizes, the entanglement entropy appears to saturate with $N$ for annealing paths crossing the first-order transition, shows sublinear scaling in the vicinity of the critical point and increases more slowly but apparently over a larger range of system sizes across the second-order transition.
    }
    \label{fig:p-spin_1}
\end{figure*}

\subsection{Quantum resources across the annealing parameter space}\label{subsec:p-spin_parameterspace}

We start by analyzing the behavior of the $p$-spin system in the $(s,\lambda)$-parameter space (Eq.~\eqref{eq:H_nonstoquastic_pspin}). The first row of Fig.~\ref{fig:p-spin_1}, corresponding to panels (a)–(c), shows, from left to right, the energy gap, the bipartite entanglement, and the SRE for a system with $N=30$ and $p=5$. The parameters $s$ and $\lambda$ are each discretized uniformly in the interval $[0,1]$, using  $301 \times 301$ grid points in total. A clear qualitative similarity emerges among the three observables: the region where the energy gap becomes minimal (white dashed line) coincides with the region where both entanglement and magic are maximized, indicating that quantum correlations and non-stabilizer resources are maximal in the critical region of the annealing process.

This behavior is consistent with the phase diagram of the $p$-spin model under non-stoquastic annealing (see Fig.~3 in Ref.~\cite{Seki2012}). The minimum energy gap, as well as entanglement and SRE peaks, occur at the transition from paramagnetic to ferromagnetic phase in the $(s,\lambda)$-plane ~\cite{Seki2012}.
For large values of $\lambda$, this boundary corresponds to a first-order transition. As $\lambda$ decreases, it terminates at a critical point around $(s,\lambda)\approx(0.4,0.3)$, beyond which the transition becomes second-order and the exponentially small gap is avoided.

Away from the line of the minimal energy gap, entanglement and SRE display qualitatively different behaviors. 
For small $s$, both remain close to zero across all $\lambda$. 
At around the critical $s$ value for each $\lambda$, both quantities reach a maximum. Afterwards, the stabilizer Rényi entropy decreases smoothly, whereas entanglement exhibits a sharper drop. In both cases, however, this variation becomes progressively smoother when moving towards the second-order transition region ($\lambda \lesssim 0.3$). 
Further, for $\lambda \gtrsim 0.3$, the entanglement entropy develops a pronounced local minimum at values of $s$ above the critical line, while the SRE instead decreases smoothly without exhibiting a corresponding minimum beyond the critical region. These differences highlight that, while both quantities respond to the same critical region, they encode distinct aspects of the underlying quantum structure of the state.

\subsection{Finite-size scaling along sinusoidal annealing paths}\label{subsec:p-spin_scaling}

Guided by the structure of the phase diagram discussed above, we consider annealing paths in which $\lambda$ is varied as a function of $s$. In particular, we adopt a sinusoidal schedule: 
\begin{equation}\label{eq:pspin_schedule}
\lambda(s) = 1- A \sin(\pi s),
\end{equation}
where the amplitude $A$ controls the strength of the non-stoquastic catalyst term during the annealing protocol (see Eq.~\eqref{eq:H_nonstoquastic_pspin}). This choice, illustrated in Fig.~\ref{fig:schedule_sketch}, serves as a representative example; more general schedules could also be considered, provided they interpolate between $s=0$ and $s=1$ and ensure $\lambda(s=1)=1$.

We consider four different values of the amplitude, $A\in \left\{0,0.5,0.7,0.9\right\}$. 
For $A=0$, the annealing path reduces to the standard stoquastic schedule; for $A=0.5$, it still crosses the line of first-order transitions; for $A=0.7$, it approaches the critical point where the transition changes from first to second order; and for $A=0.9$, the path lies entirely in the second-order region, thus avoiding first-order transitions.
The second row of Fig.~\ref{fig:p-spin_1} (panels (d)–(f)) summarizes the finite-size behavior of the three quantities along these four annealing paths. From left to right, the panels show the minimum energy gap, the maximum bipartite entanglement, and the maximum SRE attained along the annealing path, as a function of the system size.
For the energy gap and the SRE, we consider system sizes from $N=10$ to $N=60$ in steps of five, while for the entanglement entropy we extend the analysis up to $N=100$ for the non-stoquastic paths with $A=0.7$ and $A=0.9$, to better resolve their large-$N$ behavior.

\begin{figure*}[ht]
    \centering
    \includegraphics[width=1\linewidth]{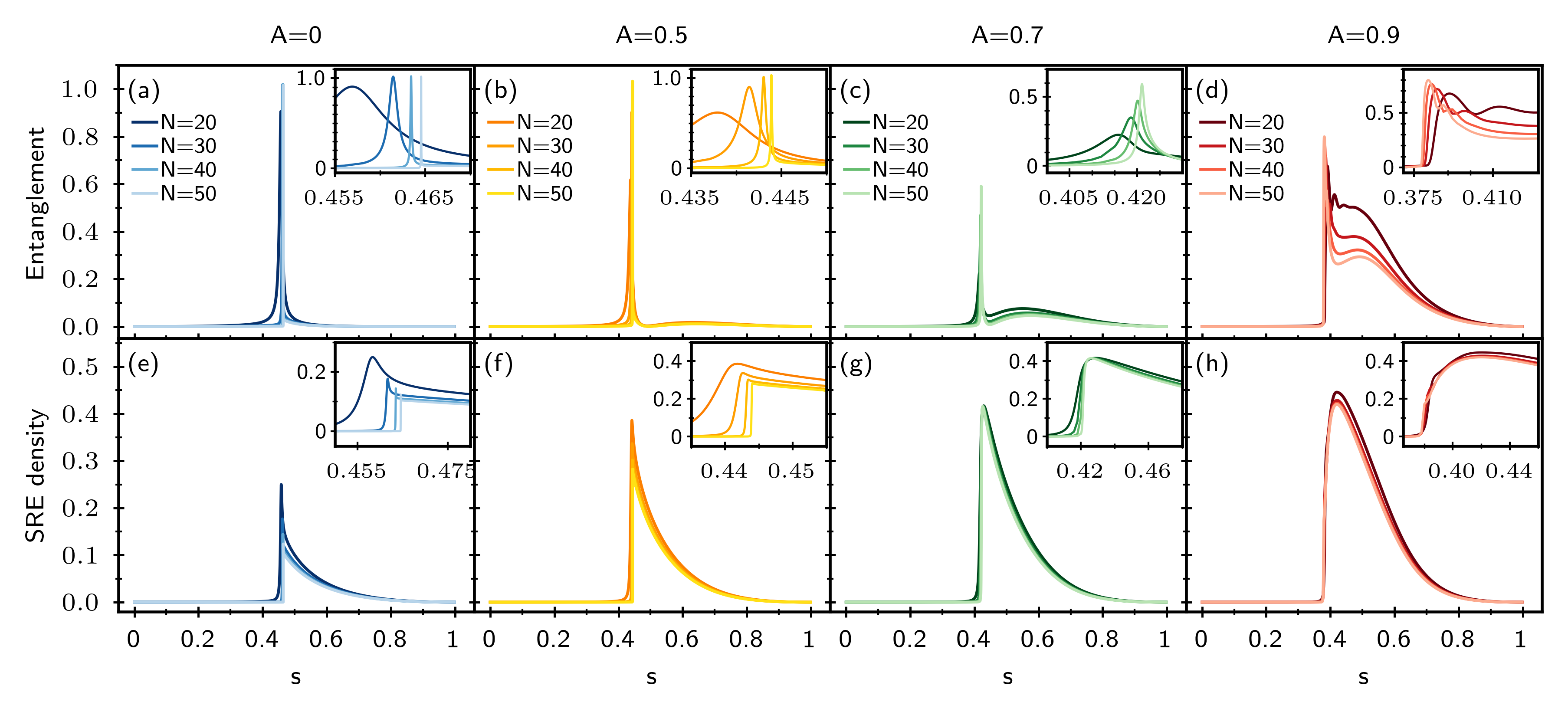}
    \caption{\textbf{Top row}: bipartite entanglement entropy of the instantaneous ground state along the annealing sweep as a function of $s$, for annealing paths defined by $\lambda(s)=1-A\sin(\pi s)$ with amplitudes $A=0,\,0.5,\,0.7,\,0.9$ (from left to right). For each amplitude, results are shown for system sizes $N=20,30,40,50$. \textbf{Bottom row}: corresponding stabilizer Rényi entropy density. Insets show the zoomed-in regions around the maximal entanglement and SRE values.
    With increasing non-stoquastic amplitude $A$, the entanglement barrier becomes broader and grows with system size. The SRE density instead changes from a sharp peak, whose height decreases with $N$ for $A=0$, $0.5$, to a broader and nearly size-independent profile for $A=0.7$, $0.9$, consistent with the extensive scaling of the total SRE.}
    \label{fig:p-spin-2}
\end{figure*}

In our numerics, we uniformly discretize the annealing parameter $s$, with an observable- and path-dependent resolution. To obtain accurate values for the minimal energy gap, we used up to $\Delta s \sim 10^{-7}$. For the stabilizer Rényi entropy and the entanglement entropy, $\Delta s \sim 10^{-6}$–$10^{-5}$ proved sufficient for smaller sizes while we employed finer resolutions up to $\Delta s \sim 10^{-7}$ for larger system sizes to obtain accurate estimates of their global maxima. Although small deviations may occur for specific data points, we found these resolutions sufficient to reliably capture the qualitative behavior (see Appendix~\ref{app:p-spin} for more details).

The minimum gap exhibits two qualitatively distinct scaling behaviors, depending on the parameter regime. This is shown via the dashed lines in Fig.~\ref{fig:p-spin_1}(d), which are obtained from polynomial fits on a logarithmic scale that retain only the leading contributions. For $A=0$ and $A=0.5$, corresponding to paths crossing the line of first-order transitions, the gap decreases exponentially with system size, as expected.  
By contrast, for $A=0.7$ and $A=0.9$, corresponding to paths that approach or lie within the second-order region of the phase diagram, the decay becomes significantly slower and is well described in logarithmic scale by a quadratic dependence on $N$, $\log \Delta E_{\min} \approx a + bN + cN^2$. This change in scaling reflects the softening of the critical bottleneck (closing of the energy gap) when first-order transitions are avoided, as previously shown in Ref.~\cite{Seki2012}.

Our main findings concern the behavior of the quantum resource measures. For the entanglement entropy, shown in Fig.~\ref{fig:p-spin_1}(e), we observe different behavior along various annealing paths. 
For $A=0$ and $A=0.5$, the maximum entanglement grows rapidly at small system sizes and then saturates to a value close to $1$ as $N$ increases further.
In contrast, for $A=0.7$, corresponding to a path close to the critical point, the maximum entanglement starts at a smaller value at small system sizes but continues to increase over the entire range of system sizes considered, suggesting stronger quantum correlations. For $A=0.9$, the growth is slower. The available system sizes do not allow us to unambiguously identify whether the growth continues in the thermodynamic limit or whether the entanglement entropy will converge. 
In any case, the benign scaling of entanglement throughout the phase diagram suggests that tensor networks, which have been applied to the stoquastic limit of the p-spin model \cite{Lami2023Thesis}, can simulate the model in the non-stoquastic regime for relatively large system sizes.

The behavior of the SRE is more regular than that of the entanglement entropy: as shown in Fig.~\ref{fig:p-spin_1}(f), in all cases, the maximum SRE grows approximately linearly with the system size, indicating an extensive scaling, well described by $M_2^{\max} = a + bN$.
However, the slope of this growth strongly depends on the annealing path: paths crossing a first-order transition ($A=0$ and $A=0.5$) exhibit a much weaker increase, 
whereas paths approaching or lying within the region where the second-order transition occurs ($A=0.7$ and $A=0.9$) display significantly steeper growth.

Thus, although the SRE is extensive for all considered protocols, it is strongly enhanced when the annealing path reaches deep into the non-stoquastic regime and avoids first-order transitions. Furthermore, we observe that the scaling behavior of the SRE is almost equivalent within numerical accuracy for all paths crossing the second-order phase transition. 

\subsection{Entanglement and non-stabilizerness barriers}\label{subsec:p-spin_barriers}

To further analyze the scaling behaviors discussed above, we investigate how entanglement and magic are distributed along the annealing sweep. Figure~\ref{fig:p-spin-2} shows the bipartite entanglement entropy (top row) and the SRE density (bottom row) along the sinusoidal annealing sweep for the four values of the amplitude $A$ considered above and system sizes $N=20,30,40,50$. The annealing parameter $s$ is discretized as explained in the previous section, with finer resolution near the peaks and a coarser step size $\Delta s = 0.002$ elsewhere.

For the entanglement entropy, we observe pronounced peaks for $A=0$, $0.5$, and $0.7$, whose sharpness increases both with $N$ for fixed $A$ and with decreasing $A$ for fixed $N$.
For $A=0.9$, the structure changes qualitatively: the profile develops multiple peaks and extends over a wider range of $s$. In all cases, the maximum entanglement increases with system size. 
A different trend is observed for the SRE density. As the amplitude $A$ increases, the profile becomes broader and smoother. For $A=0$ and $A=0.5$, the peak height decreases with system size.
In contrast, for $A=0.7$ and $A=0.9$, it remains approximately constant, consistent with the extensive scaling of the total SRE observed in the finite-size analysis outside the first-order transition region.

Overall, the two resources respond differently to the non-stoquastic catalyst. The maximum entanglement entropy initially grows with system size and approaches saturation for larger $N$. However, the rate of convergence becomes slower for strongly non-stoquastic paths ($A=0.7,0.9$). This suggests that tensor-network methods may remain effective in large parts of the parameter regime, although such compatibility is not guaranteed for generic non-stoquastic paths. By contrast, the stabilizer Rényi entropy grows extensively with $N$ in all cases, with an increasing slope for stronger catalysts. Thus, the regime where non-stoquasticity improves the minimum-gap scaling also corresponds to the regime generating the largest amount of non-stabilizerness, hindering efficient Clifford-based classical simulation.

\section{Local Ising model}\label{sec:localising}

As a second benchmark problem, we consider a geometrically local Ising model for which a potential advantage of non-stoquastic catalysts has been reported~\cite{Albash2019}. Unlike the $p$-spin model, this system lacks permutation symmetry, involves only two-body interactions, and features geometrically local qubit connectivity, making it more relevant for direct experimental implementations. In this section, we introduce the model and its discrete symmetries (Sec.~\ref{subsec:local_ising_definition}), analyze the magnetization landscape induced by the catalyst (Sec.~\ref{subsec:local_ising_magnetization}), 
and finally study the energy gap, bipartite entanglement, and stabilizer Rényi entropy across the annealing parameter space (Sec.~\ref{subsec:local_ising_parameterspace}) and their finite-size scaling (Sec.~\ref{subsec:local_ising_scaling}).

\subsection{Problem definition and symmetries}
\label{subsec:local_ising_definition}

\begin{figure}
    \centering
    \includegraphics[width=\linewidth]{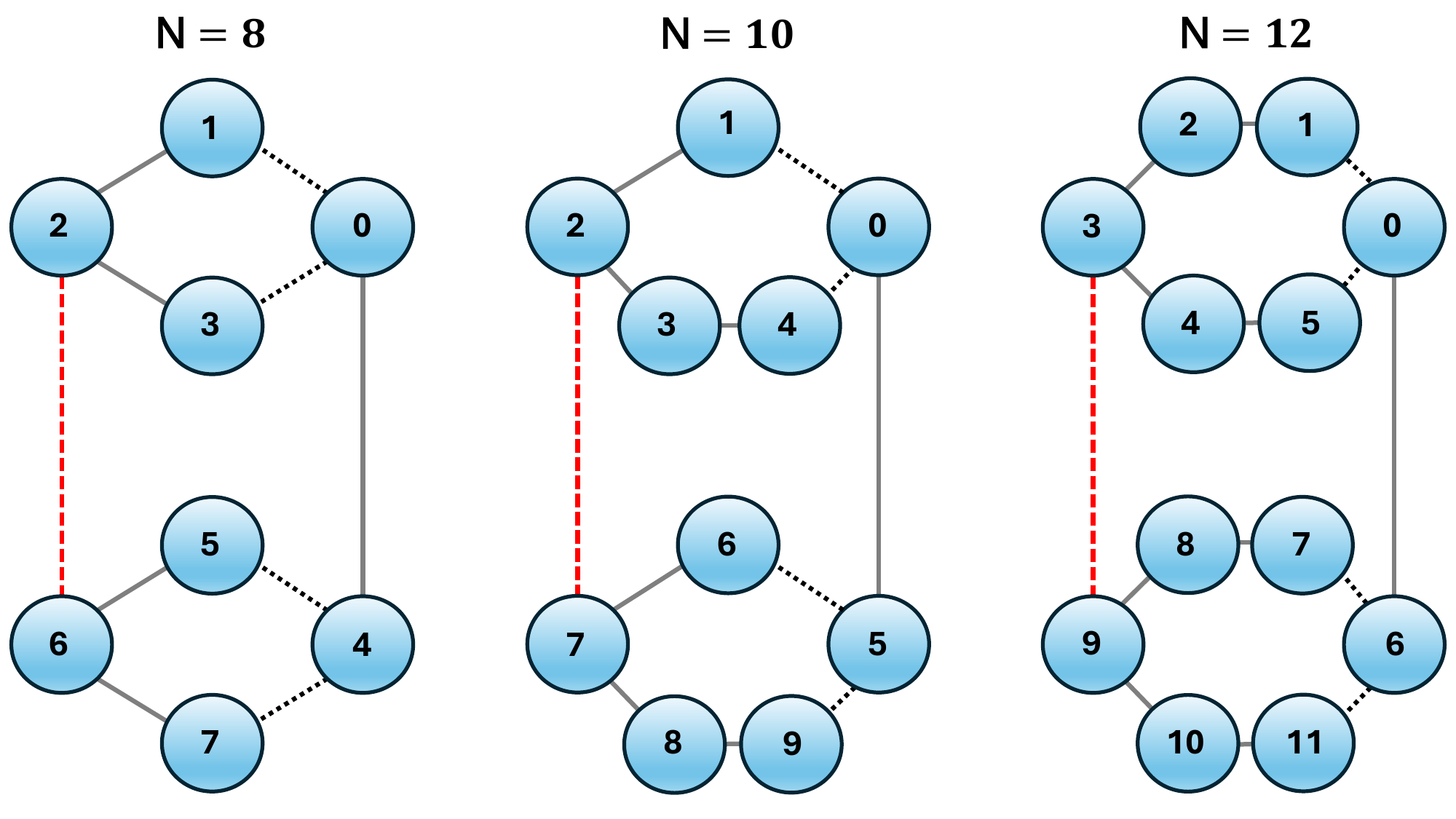}
    \caption{Graph structure of the cost Hamiltonian $H_\mathrm{cost}$ in the local Ising model defined by Eq.~\eqref{eq:H_cost_albash} for system sizes $N=8$, $10$, and $12$ (introduced in Ref.~\cite{Albash2019}). Each graph consists of two rings of $N/2$ qubits connected by two inter-ring couplings at diametrically opposite positions. Solid gray lines denote ferromagnetic couplings of strength $J_{ij} = -1$, dashed black lines ferromagnetic couplings of strength $J_{ij} = -1/2$, and red dashed lines antiferromagnetic couplings of strength $J_{ij} = +5/6$. The catalyst Hamiltonian $H_\mathrm{catalyst}$, see Eq.~\eqref{eq:H_cat_albash}, shares the same connectivity but with uniform antiferromagnetic couplings. The site labels define the numbering convention used
    in our numerics.}
    \label{fig:local_ising_sketch}
\end{figure}

The annealing Hamiltonian of the geometrically local Ising model takes the form
\begin{equation}
\label{eq:H_albash}
\hat{H}(s, \lambda) = (1-s)\,H_\mathrm{drive} + s\,H_\mathrm{cost} + \lambda\, s(1-s)\,H_\mathrm{catalyst} \,,
\end{equation}
with
\begin{align}
H_\mathrm{drive} &= -\sum_i \sigma_i^x \,, \\
H_\mathrm{cost} &= \sum_{\langle i,j \rangle} J_{ij}\, \sigma_i^z \sigma_j^z \,, \label{eq:H_cost_albash} \\
H_\mathrm{catalyst} &= \sum_{\langle i,j \rangle} \sigma_i^x \sigma_j^x \,, \label{eq:H_cat_albash}
\end{align}
where the sums run over nearest-neighbor pairs $\langle i,j \rangle$ on a graph consisting of two rings of $N/2$ qubits each, with nearest-neighbor couplings along each ring. The rings are connected by two inter-ring couplings linking diametrically opposite sites (see Fig.~\ref{fig:local_ising_sketch} and Ref.~\cite{Albash2019}). 
The annealing schedule in Eq.~\eqref{eq:H_albash} differs from that of the $p$-spin model: following the convention in Ref.~\cite{Albash2019}, $\lambda$ is a fixed parameter controlling the catalyst strength, while the envelope $s(1-s)$ ensures that the catalyst is active only at intermediate times. By contrast, in the $p$-spin model Hamiltonian both $s$ and $\lambda$ are time-dependent.

The catalyst Hamiltonian shares the same connectivity as the cost Hamiltonian, and for $\lambda > 0$ introduces positive off-diagonal matrix elements rendering the total Hamiltonian non-stoquastic. For the stoquastic case ($\lambda \leq 0$), the minimum energy gap closes exponentially with system size due to a perturbative level crossing near $s = 1$~\cite{Albash2019}. In contrast, for suitable positive $\lambda$ the catalyst splits this crossing into multiple avoided crossings, leading to a significantly improved, potentially polynomial scaling of the minimum gap.

\textit{Numerical methods.}--- 
The Hamiltonian~\eqref{eq:H_albash} possesses two discrete symmetries that enable a constant-factor reduction of the Hilbert space: the parity (spin-flip) symmetry $\hat{\Pi} = \prod_{i=1}^N \sigma_i^x$ and the ring-swap symmetry $\hat{P}_\mathrm{swap}$, which exchanges the two rings (see Fig.~\ref{fig:local_ising_sketch}). Both commute with the Hamiltonian and with each other, generating a $\mathbb{Z}_2 \times \mathbb{Z}_2$ symmetry group.
Since the initial state $\ket{+}^{\otimes N}$ belongs to the joint $+1$ eigenspace of both operators and the symmetries are preserved during the evolution, the dynamics is restricted to this symmetric sector. Its dimension scales as $d \sim 2^N/4$ for large $N$, providing a constant-factor reduction while retaining an overall exponential scaling.
Within this reduced space, we perform exact diagonalization to obtain the instantaneous eigenstates and eigenvalues, from which we compute the energy gap, the entanglement entropy and the SRE. The latter two quantities are evaluated after reconstructing the wavefunction in the full computational basis. The bipartite entanglement, whose calculation scales as $\mathcal{O}(2^{\frac{3N}{2}})$, can still be computed exactly for moderate system sizes. In contrast, the evaluation of the SRE becomes quickly intractable due to the exponential number of Pauli strings. To access larger systems, we therefore employ matrix product state (MPS) representations combined with Monte Carlo sampling over Pauli operators~\cite{Tarabunga2023}, at the cost of controlled truncation and statistical errors. In our implementation, the MPS bond dimension is set to $\chi = 16$ for $N \leq 14$ and $\chi = 32$ for $N \geq 16$. The SRE is estimated by averaging over $10$ independent Monte Carlo runs, with $10^4$ samples for $N < 16$ and $3 \times 10^4$ samples for larger systems (see Appendix~\ref{app:local-ising}).

\begin{figure*}
    \centering
     \includegraphics[width=1\linewidth]{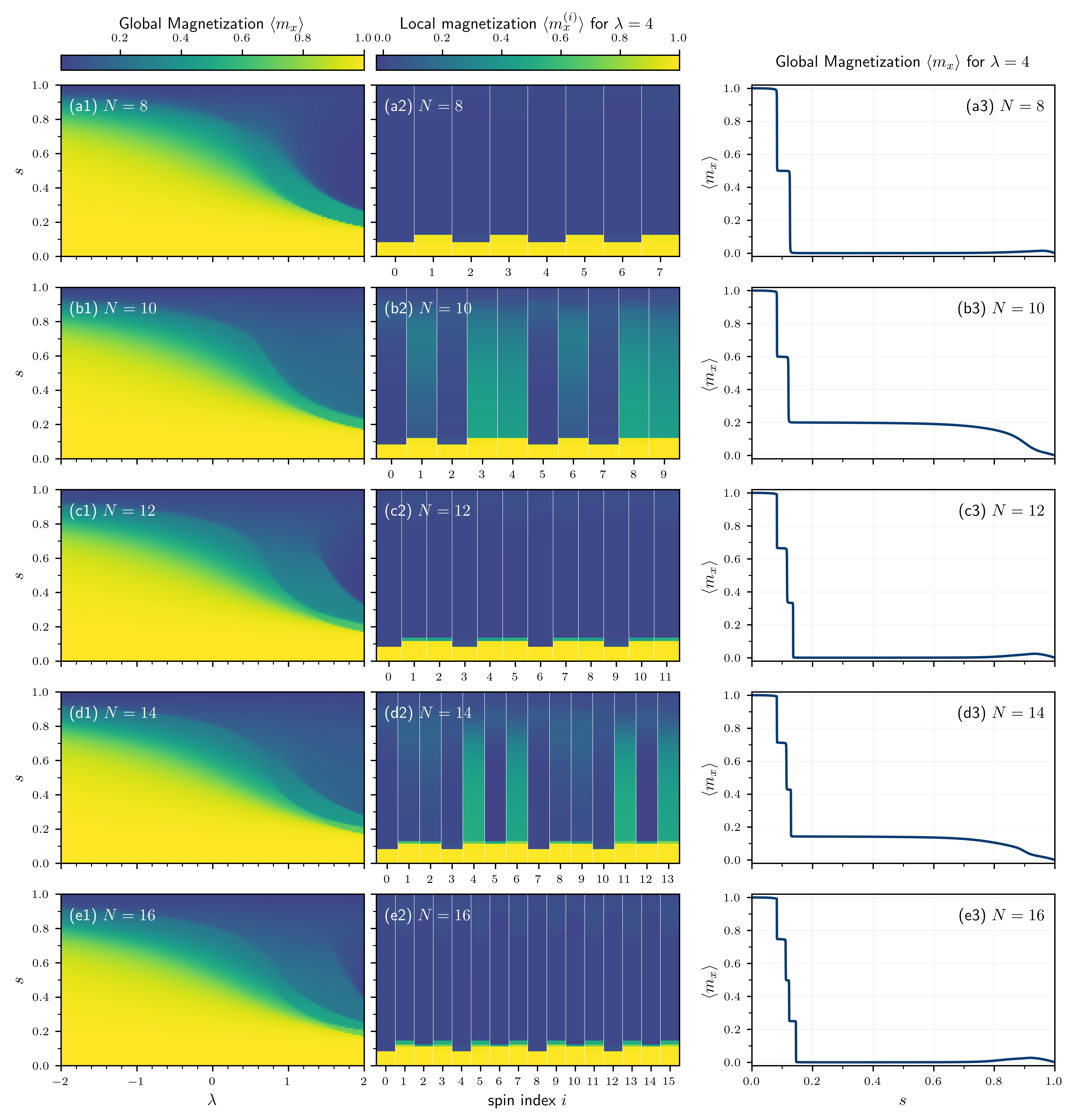}
    \caption{Magnetization across Hamiltonian parameter space $(\lambda,s)$ and spin-resolved dynamics of the local Ising model (Eq.~\eqref{eq:H_albash}) for system sizes $N=8$--$16$ (rows). 
\textbf{Left column:} Average transverse magnetization in the parameter plane spanned by the annealing parameter $s$ and the non-stoquasticity $\lambda$, discretized into a 401x101 grid. 
\textbf{Middle column:} Spin-resolved local magnetization $\langle m_x^{(i)} \rangle$ versus annealing time $s$ for fixed $\lambda=4$.   
\textbf{Right column:} Vertical annealing sweeps at fixed $\lambda=4$, highlighting step-like plateaus in $\langle m_x \rangle$ and abrupt crossovers between magnetization sectors as $s$ varies. The non-stoquastic catalyst ($\lambda > 0$) replaces the single magnetization
rearrangement of the stoquastic regime with a sequence of plateaus, each corresponding to a distinct competing spin configuration, that sharpen with increasing $\lambda$.}
    \label{fig:magnetization}
\end{figure*}

\subsection{Transitions in the magnetization landscape}
\label{subsec:local_ising_magnetization}

To characterize the behavior of the system along the annealing path, we study the magnetization as a function of the interpolation parameter $s$ and the catalyst strength~$\lambda$. Throughout this section, $\ket{\psi_0(s,\lambda)}$ denotes the instantaneous ground state of the Hamiltonian in Eq.~\eqref{eq:H_albash}. The mean global transverse magnetization is defined as
\begin{equation}
\label{eq:mx_global}
\langle m_x \rangle = \frac{1}{N} \sum_{i=1}^{N} \bra{\psi_0} \sigma_i^x \ket{\psi_0} \,,
\end{equation}
which equals unity at $s = 0$ (where the ground state is $\ket{+}^{\otimes N}$) and vanishes at $s = 1$ (where the ground state is an eigenstate of $H_\mathrm{cost}$). 
Its behavior illustrates how the ground state reorganizes along the annealing path: sharp variations in the thermodynamic limit are indicative of a first-order transition, while smooth variations are characteristic of a second-order transition. 
In the following, we use this observable as our primary diagnostic tool to characterize the evolution of the ground state along the annealing sweep and to identify the different regimes induced by the catalyst.

In Fig.~\ref{fig:magnetization}, the first column (panels (a1)–(e1)) shows heatmaps of the transverse magnetization $\langle m_x \rangle$ as a function of $s$ and $\lambda$ for system sizes $N = 8, 10, 12, 14,16$. 
For negative values of $\lambda$, the magnetization varies smoothly along the annealing path until the final stage, where the system quickly approaches the ground state with vanishing transverse magnetization. In contrast, for positive $\lambda$ a significantly richer structure is observed.
In this regime, regions of rapid variation emerge in the $(s,\lambda)$ plane and sharpen with increasing system size. 
For large $\lambda$, where the structure becomes increasingly regular, the magnetization shows a sequence of plateaus at discrete values determined by the allowed configurations, whose number increases with $N$.

For finite-size systems, the magnetization profile remains continuous: the apparent plateaus are connected by narrow but smooth rearrangements of the ground state. These rearrangements originate from a sequence of avoided level crossings between competing low-energy configurations, as analyzed in detail in Appendix~\ref{app:local-ising-effective-model}. The third column of Fig.~\ref{fig:magnetization} (panels (a3)–(e3)) shows the global transverse magnetization along a vertical sweep at fixed $\lambda = 4$, making the plateau structure and the steep transitions between plateaus directly visible. The spatial structure of the plateau states is resolved by the local transverse magnetization,
\begin{equation}
\label{eq:mx_local}
\langle m_x^{(i)} \rangle = \bra{\psi_0} \sigma_i^x \ket{\psi_0} \,,
\end{equation}
shown for $\lambda = 4$ in the second column of Fig.~\ref{fig:magnetization} (panels (a2)–(e2)). Because the model lacks permutation symmetry, $\langle m_x^{(i)} \rangle$ varies across lattice sites, and each plateau is characterized by a distinct pattern in which specific sites remain fully polarized while others become depolarized. As discussed in Appendix~\ref{app:local-ising-effective-model}, these patterns correspond to the spin configurations that minimize the energy of an effective large-$\lambda$, small-$s$ Hamiltonian, where the cost term contributes only perturbatively.

We derive this effective description by expanding Eq.~\eqref{eq:H_albash} to leading
order in $s$. Taking $\lambda\gg 1$ and $s\ll 1$, but $\tilde{\lambda} \equiv \lambda s$ fixed, and using
$s(1-s) \approx s$, we obtain
\begin{equation}
\label{eq:H_effective_smalls}
\hat{H}_\mathrm{eff} = -\sum_i \sigma_i^x
    + \tilde{\lambda} \sum_{\langle i,j \rangle} \sigma_i^x \sigma_j^x
    + \mathcal{O}(s) \,,
\end{equation}
a longitudinal-field Ising model with competing on-site magnetic field and nearest-neighbor interactions in which the cost Hamiltonian drops out. All terms in Eq.~\eqref{eq:H_effective_smalls} commute, so the ground state is determined by minimizing a classical energy over spin configurations in the $\sigma^x$ basis. 
When $\tilde{\lambda} \ll 1$, the longitudinal field dominates and the ground state is fully polarized. For $\tilde{\lambda} \gg 1$, the longitudinal field becomes negligible and the system is dominated by the $\sigma^x \sigma^x$ interactions, so that a discrete sequence of competing spin configurations becomes successively favored.
Their classical level crossings determine the plateau boundaries observed in Fig.~\ref{fig:magnetization}. 
The parity symmetry $\hat{\Pi}$ constrains the dynamics to the even-parity sector, so that successive plateaus differ by an even number of flipped spins; the explicit configurations, plateau energies, and crossing points are worked out in Appendix~\ref{app:local-ising-effective-model}. Contributions subleading in $s$ introduce non-commuting terms to Eq.~\eqref{eq:H_effective_smalls}, which smoothen the sharp transitions between successive plateaus. 
 
This effective description clarifies the role of the non-stoquastic catalyst. Already at small $s$, the $\sigma^x \sigma^x$ term drives the ground state through a sequence of spin rearrangements with progressively smaller global magnetization rather than a single transition. We emphasize that these intermediate rearrangements do not directly enable improved annealing performance, which instead arises from the interplay between cost and catalyst Hamiltonians. The optimal regime for quantum annealing was identified in Ref.~\cite{Albash2019} to lie around $\lambda \approx 1.5$, with the exact value being strongly dependent on system size.

\subsection{Quantum resources across the annealing parameter space}
\label{subsec:local_ising_parameterspace}

\begin{figure*}
    \centering
    \includegraphics[width=1\linewidth]{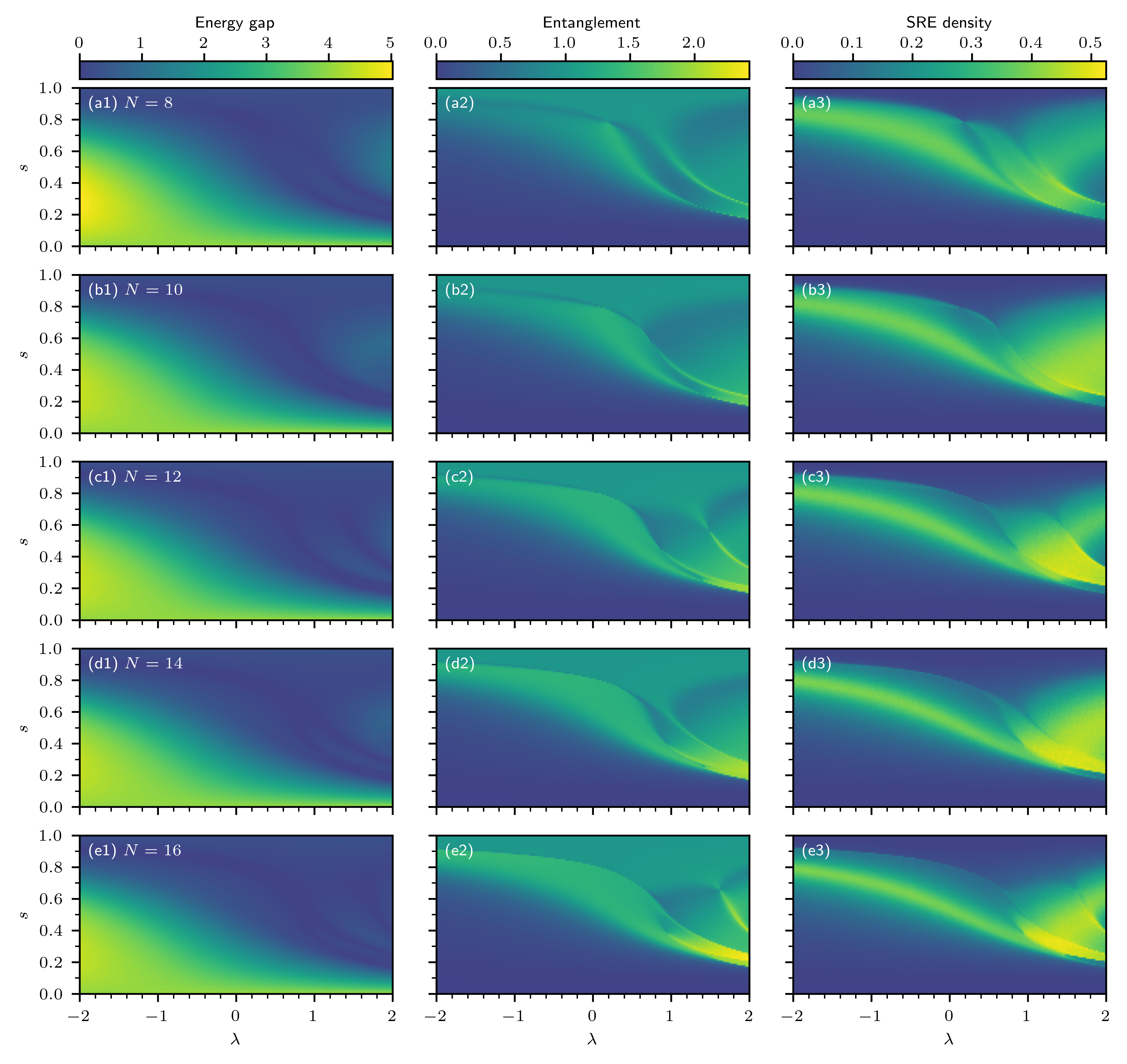}
    \caption{Energy gap and quantum resources evaluated on the instantaneous ground state of the Hamiltonian in Eq.~\eqref{eq:H_albash}, as a function of the annealing parameters $s$ and $\lambda$, and for system sizes $N = 8, 10, 12, 14, 16$ (from top to bottom). 
    \textbf{First column}: Energy gap. \textbf{Middle column}: Bipartite entanglement entropy. \textbf{Third column}: Stabilizer R\'enyi entropy density. The non-stoquastic catalyst ($\lambda > 0$) replaces the single transition of the stoquastic protocol with a sequence of gap minima, each coinciding with local maxima or the onset of plateaus of both quantum resources.}
    \label{fig:local-ising-grids}
\end{figure*}

As in the case of the $p$-spin model (Sec.~\ref{sec:pspin}), we characterize the system through three observables evaluated for the instantaneous ground state: the energy gap, the bipartite entanglement entropy, and the SRE. Details on the numerical methods used to compute these quantities are provided in Appendix~\ref{app:local-ising}.
We analyze these quantities as a function of the $(\lambda, s)$ parameters of the annealing Hamiltonian (Eq.~\eqref{eq:H_albash}), where $\lambda \in [-2,2]$ controls the catalyst strength and $s\in[0,1]$ parametrizes the annealing evolution. The parameter space is discretized on a $401 \times 101$ grid in $(\lambda,s)$. Figure~\ref{fig:local-ising-grids} shows the resulting landscapes for system sizes $N=8,10,12,14,16$ (from top to bottom): the first column (panels (a1)–(e1)) reports the energy gap, the second column (panels (a2)–(e2)) the bipartite entanglement entropy, and the third column (panels (a3)–(e3)) the SRE density.

For $\lambda \leq 0$, the three quantities display local extrema around a single line in the $(\lambda,s)$ plane. Along this line, the energy gap together with the entanglement entropy develops a pronounced minimum, while the SRE develops a corresponding local maximum. All quantities exhibit sharp features that become increasingly narrow as the system size grows. This behavior indicates that in the stoquastic regime the annealing dynamics is dominated by a single rearrangement of the ground state.

For $\lambda > 0$, the structure of the annealing landscape changes qualitatively and matches the behavior observed in the transverse magnetization (see Fig.~\ref{fig:magnetization}). In this regime, the observables display multiple features: the energy gap develops several local minima, accompanied by corresponding local maxima or plateaus of entanglement and SRE (for more details, see Appendix~\ref{app:local_ising_barriers}). This reflects the presence of multiple competing rearrangements of the ground state induced by the non-stoquastic catalyst as discussed above and as further detailed in Appendix~\ref{app:local-ising-effective-model}.

\begin{figure*}
    \centering
    \includegraphics[width=1\linewidth]{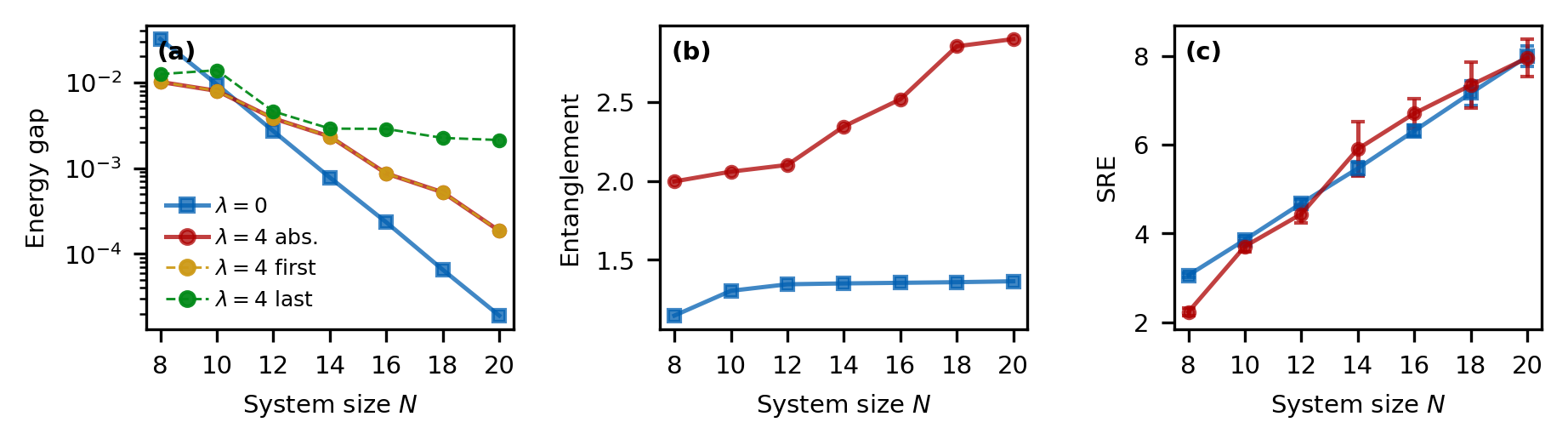}
    \caption{Finite-size scaling of (a) the minimum energy gap, (b) the maximum bipartite entanglement entropy, and (c) the maximum stabilizer Rényi entropy along annealing protocols at fixed values of $\lambda$. Results are shown for the stoquastic case $\lambda=0$ (blue squares, solid lines) and the non-stoquastic case $\lambda=4$. In the latter case, multiple local minima of the energy gap appear along the annealing path. We report the absolute minimum (red dots, solid lines), as well as the first (yellow, dashed lines) and last (green, dashed lines) local minima. For the entanglement entropy and SRE, whose profiles exhibit a more complicated structure without clearly separable peaks and plateaus, we report only the maximum value attained along the path. Lines are intended as guides to the eye. Compared with the stoquastic protocol, the non-stoquastic catalyst slows the closing of the minimum energy gap and enhances the entanglement entropy, while the SRE grows extensively with system size at approximately the same rate in both cases.}
    \label{fig:local-ising-scaling}
\end{figure*}

\subsection{Finite-size scaling along annealing protocols}
\label{subsec:local_ising_scaling}

To further investigate this behavior, we consider annealing protocols at fixed values of $\lambda$ and analyze the finite-size scaling of the quantum resources. For each protocol, we compute along the annealing sweep ($s\in[0,1]$) the minimum energy gap and the maximum values of both bipartite entanglement and SRE. We focus on two representative cases: the stoquastic protocol with $\lambda = 0$, and a non-stoquastic protocol with $\lambda = 4$. The resulting scaling behavior as a function of $N$ is reported in Fig.~\ref{fig:local-ising-scaling}.

The annealing parameter $s$ is discretized non-uniformly, with resolutions chosen to accurately resolve the relevant extrema. For the stoquastic case ($\lambda=0$), a uniform step size $\Delta s = 10^{-4}$ is sufficient for all computed quantities, while for the non-stoquastic protocol ($\lambda=4$) finer resolutions down to $\Delta s = 10^{-5}$ are used near the extrema. More details are provided in Appendix~\ref{app:local-ising}.

To account for the multiple local extrema observed at large $\lambda$, we distinguish between different relevant extrema along the evolution. In particular, for the non-stoquastic case we consider both the global minimum and the first and last occurring local minima of the energy gap. For the entanglement and SRE, we only report the global maximum, as the structure is more complex and local extrema cannot be easily identified (see Appendix~\ref{app:local_ising_barriers}).

For the stoquastic protocol ($\lambda=0$), our data suggest an exponential closing of the minimum gap with system size, consistent with Ref.~\cite{Albash2019}. In this regime, the entanglement entropy remains approximately constant, while the stabilizer Rényi entropy grows roughly linearly with $N$, indicating an extensive generation of the non-stabilizerness.

For the non-stoquastic protocol, the behavior is more complex. Although the optimal value associated with the largest minimum gap at finite system sizes was identified around $\lambda\approx 1.5$ in Ref.~\cite{Albash2019}, we focus here on $\lambda=4$. For smaller values of $\lambda$, the stronger competition between the cost Hamiltonian and the catalyst term produces a more intricate energy landscape, making the scaling behavior harder to identify. In contrast, $\lambda=4$ yields more regular structures, allowing for a more systematic analysis (see Appendix~\ref{app:local_ising_barriers}). Both the first and the last local minima in the energy gap show a slower decrease compared to the stoquastic case. 
Although the precise functional form cannot be unambiguously determined from the accessible system sizes, the available data suggest an improvement in the scaling and thus in the possible annealing speed. The entanglement entropy exhibits larger values and a steeper growth with system size compared to the stoquastic case, reflecting enhanced quantum correlations induced by the catalyst. 
For the stabilizer Rényi entropy, the identification of local extrema along the annealing path is less clear due to the intricate structure of the annealing landscape and the numerical approximations involved in the calculation. Nevertheless, when considering the global maximum along the annealing path, the SRE grows approximately linearly with $N$, indicating an overall extensive scaling also in the non-stoquastic case. 

Overall, these finite-size results suggest that the non-stoquastic catalyst modifies the scaling of the energy gap and gives rise to a richer structure in the generation of quantum resources along the annealing path.

\section{Conclusions}\label{sec:conclusion}

In summary, we have investigated non-stoquastic quantum annealing by monitoring, along the annealing sweep, the spectral gap and two complementary quantum resources of the instantaneous ground state, namely the bipartite entanglement and the non-stabilizerness. We have quantified these resources using von Neumann entropy and second stabilizer Rényi entropy, respectively. 
If the system contains an extensive amount of either resource, classical simulations based on tensor networks or stabilizer tableaus, respectively, become exponentially hard. We have analyzed two benchmark problems: (i) the fully connected ferromagnetic $p$-spin model, for which an antiferromagnetic transverse catalyst is known to turn a first-order into a second-order transition, resulting in an exponential speedup of the annealing sweep~\cite{Seki2012}, and (ii) a geometrically local Ising model~\cite{Albash2019}, where a similar advantage has been conjectured based on numerical simulations for small system sizes.

For the $p$-spin model, all three quantities---gap, entanglement, non-stabilizerness---display a qualitatively consistent picture across the $(s,\lambda)$ plane for a fixed system size. Both the entanglement entropy and the stabilizer Rényi entropy exhibit global peaks at the region of the minimum energy gap, which tracks the paramagnetic to ferromagnetic transition. 

The finite-size scaling reveals that the peaks of the two resources display different behaviors. 
The maximum entanglement rapidly saturates for stoquastic and weakly non-stoquastic protocols crossing the first-order line, while paths through the second-order transition display a slower growth reaching moderate large entanglement values with system size. By contrast, the stabilizer R\'enyi entropy grows linearly with system size for all protocols considered, with a slope that increases with the non-stoquastic strength and appears to remain unchanged near the first-to-second-order crossover.

In the geometrically local Ising model, we have observed a qualitatively different scenario, characterized by multiple competing avoided crossings and an associated complex landscape of quantum resources. As shown by the transverse magnetization, non-stoquastic catalysts do not simply modify the single dominant phase transition present in the stoquastic case, but instead induce a sequence of spin rearrangements between competing configurations. This leads to multiple local minima of the energy gap and corresponding features in entanglement and magic. While the gap appears to close more slowly and entanglement is generally enhanced in the non-stoquastic protocol, the stabilizer Rényi entropy grows extensively in both cases, and no clear asymptotic distinction can be established within the accessible system sizes, suggesting significant small-size effects.

Overall, we find that non-stoquastic annealing protocols exhibit larger or equal extensive scaling of the maximum SRE with system size compared to the stoquastic protocol.
This behavior, observed across both benchmark models, indicates the generation of strong non-Clifford resources, hindering efficient stabilizer-based classical simulation. 
The behavior of the entanglement entropy is instead less clear and appears to depend more sensitively on the annealing path and on the underlying model. In the $p$-spin model, the entanglement entropy grows slowly with system size for non-stoquastic protocols approaching the second-order transition regime, but the accessible sizes do not allow us to determine its asymptotic scaling. In the local Ising model, the entanglement entropy is generally enhanced by the non-stoquastic catalyst and continues to increase within the accessible system sizes, although no clear scaling prediction can be extracted. These results suggest that the role of tensor-network simulability in non-stoquastic annealing remains more subtle and model-dependent.

A natural direction for future work will be to better understand the scaling behavior of entanglement in non-stoquastic optimization protocols, and to determine whether common features emerge across different models and optimization strategies. To understand whether common resource-generation mechanisms emerge more generally across quantum optimization algorithms, it will also be interesting to investigate the interplay between non-stoquasticity and quantum resources beyond the adiabatic regime, including diabatic annealing protocols \cite{Takahashi2017} and variational algorithms such as the Quantum Approximate Optimization Algorithm (QAOA)~\cite{Farhi2014, Blekos2024}.

\begin{acknowledgments}
We thank Javed Akram, Marcel Seelbach Benkner, Sebastian Rubbert, and Michael Johanning for useful discussions. This project received funding by the German Federal Ministry of Research, Technology and Space under the funding reference number 13N16437 (project MAGIC App in collaboration with eleQtron GmbH). This work has benefited from Q@TN, the joint lab between University of Trento, FBK—Fondazione Bruno Kessler, INFN—National Institute for Nuclear Physics, and CNR—National Research Council. We acknowledge support by Provincia Autonoma di Trento. 
\end{acknowledgments}

\bibliographystyle{apsrev4-2}
\bibliography{trappedions}

\appendix
\section{Numerical methods for p-spin model}
\label{app:p-spin}

This appendix details how permutation symmetry, and for even $p$ also spin-flip symmetry, is used to reduce the non-stoquastic $p$-spin annealing Hamiltonian to a numerically tractable Dicke-sector representation for exact diagonalization and efficient evaluation of observables.

The $p$-spin Hamiltonian of Eq.~\eqref{eq:Hcost_pspin} is permutation symmetric,
so starting from the initial state $\ket{+}^{\otimes N}$, which lies in the
fully symmetric subspace, the dynamics is confined to the $(N+1)$-dimensional
Dicke sector spanned by the states $\ket{S, m}$ with $S = N/2$ and
$m = -S, \ldots, S$. Introducing the collective spin operators
$\hat{S}^\alpha = \frac{1}{2}\sum_{i=1}^N \hat{\sigma}_i^\alpha$ for
$\alpha = x, y, z$, the cost, driver, and catalyst Hamiltonians of
Eq.~\eqref{eq:H_nonstoquastic_pspin} take the form
\begin{align}
\label{eq:Hcost-dicke}
\hat{H}_\mathrm{cost}     &= -2^p N^{1-p} (\hat{S}^z)^p \,, \\
\label{eq:Hdrive-dicke}
\hat{H}_\mathrm{drive}    &= -2\,\hat{S}^x \,, \\
\label{eq:Hcatalyst-dicke}
\hat{H}_\mathrm{catalyst} &= \frac{4}{N}\,(\hat{S}^x)^2 \,.
\end{align}
The cost term is diagonal in the Dicke basis, with matrix elements
$\bra{S,m} \hat{H}_\mathrm{cost} \ket{S,m} = -2^p N^{1-p}\, m^p$, while the
driver and catalyst are off-diagonal. Their action follows from the standard
ladder representation $\hat{S}^x = (\hat{S}^+ + \hat{S}^-)/2$ with
$\hat{S}^\pm \ket{S,m} = \sqrt{S(S+1) - m(m\pm 1)}\,\ket{S,m\pm 1}$, so
$\hat{H}_\mathrm{drive}$ couples states with quantum number $m$ to $m \pm 1$, and
$\hat{H}_\mathrm{catalyst}$ couples $m$ to $m$ and
$m \pm 2$. The full annealing Hamiltonian of Eq.~\eqref{eq:H_nonstoquastic_pspin} therefore
acts on the Dicke sector as an $(N+1)\times(N+1)$ matrix. For even $p$, the additional spin-flip symmetry
$\hat{\Pi} = \prod_i \hat{\sigma}_i^x$, which acts as $\ket{S,m} \mapsto
\ket{S,-m}$ in the Dicke basis, further reduces the relevant dimension. This reduced space has
dimension $(N+1)/2$ for odd $N$, and $N/2+1$ for even $N$, the extra state in
the latter case being the unpaired Dicke state $\ket{S,0}$. Exact diagonalization within these subspaces gives
access to system sizes of several hundred qubits; in this work, we use up to
$N = 100$. The instantaneous ground state is expanded as
\begin{equation}
\label{eq:dicke-expansion}
\ket{\psi} = \sum_{m=-S}^{S} c_m \ket{S, m} \,,
\end{equation}
from which the observables of interest are computed as follows.

\paragraph{Bipartite entanglement entropy.}
For a balanced bipartition $n_A = n_B = N/2$, the reduced density matrix
$\rho_A = \mathrm{Tr}_B(\ket{\psi}\bra{\psi})$ can be evaluated directly from
the Dicke coefficients $c_m$ without reconstructing the $2^N$-dimensional
state. Using the Schmidt decomposition of the Dicke states~\cite{Stockton2003}, one finds
$\rho_A = A A^\dagger$ with
\begin{equation}
\label{eq:Amatrix}
A_{m,j} = c_{m+j}\,\sqrt{\frac{\binom{n_A}{m}\binom{n_B}{j}}{\binom{N}{m+j}}} \,,
\end{equation}
which has dimension $(n_A+1)\times(n_B+1)$. The bipartite entanglement entropy
of Eq.~\eqref{eq:entanglement-entropy} then follows directly from the
eigenvalues of $\rho_A$.

\paragraph{Stabilizer R\'enyi entropy.}
Direct evaluation of the SRE of Eq.~\eqref{eq:SRE} requires, in general,
computing $4^N$ Pauli expectation values, which is prohibitive already for
modest $N$. For permutation-symmetric states, however, the expectation value
$\bra{\psi} P \ket{\psi}$ of a Pauli string
$P = \hat{\sigma}^{\mu_1}_{i_1}\cdots\hat{\sigma}^{\mu_N}_{i_N}$, with
$\mu_k \in \{0,x,y,z\}$ and
$\hat{\sigma}^0 \equiv 1$, depends only
on the counts $(N_X, N_Y, N_Z, N_0)$ of the four Pauli labels in $P$, not on
which qubits they act on. Pauli strings can therefore be grouped into
equivalence classes labeled by these counts, of which there are
\begin{equation}
D = \binom{N+3}{3} = \mathcal{O}(N^3) \,,
\end{equation}
with multiplicities calculated as $ N! / (N_X! N_Y! N_Z! N_0!)$. This
reduces the evaluation of the SRE in the Dicke basis from an exponential to a
polynomial cost in $N$~\cite{Passarelli2024}, enabling access to the same
system sizes as the entanglement entropy.

\paragraph{Discretization of the annealing parameter.}

The annealing parameter $s$ is discretized uniformly, with higher resolution
in the vicinity of the relevant extrema, namely the minima of the energy gap and
the maxima of the bipartite entanglement entropy and SRE. For the energy gap, we
use $\Delta s = 10^{-7}$ at $A = 0$ and $\Delta s = 5\times 10^{-6}$ at
$A = 0.5, 0.7, 0.9$. For the bipartite entanglement entropy, we use
$\Delta s = 5\times 10^{-7}$ at $A = 0$ and $\Delta s = 5\times 10^{-6}$ for
the remaining amplitudes. For the largest sizes, $N = 70, 80, 90, 100$, this
grid is further refined to $\Delta s = 10^{-7}$ at $A = 0.7$ and $A = 0.9$.
For the SRE, whose evaluation is computationally more expensive, we use a
slightly coarser grid, with $\Delta s = 5\times 10^{-6}$ at $A = 0$ and
$\Delta s = 5\times 10^{-5}$ at the remaining amplitudes. We find that these
discretizations give satisfactorily converged estimates of the extremal values
of all considered observables.

\section{Symmetries and numerical methods for the local Ising model}
\label{app:local-ising}

This appendix summarizes the symmetry reduction and numerical procedures used to obtain the numerical data for the local Ising model discussed in Sec.~\ref{sec:localising}.

For this geometrically local Ising model, the absence of
permutation symmetry rules out the same polynomial-dimensional reduction as available
for the $p$-spin model (Dicke basis). The Hamiltonian of Eq.~\eqref{eq:H_albash} nonetheless
possesses two discrete symmetries that together yield a constant-factor
reduction of the Hilbert space, which we detail in the following.

\paragraph{Discrete symmetries.}
Although Eq.~\eqref{eq:H_albash} is not permutation symmetric, it is invariant under
two commuting $\mathbb{Z}_2$ symmetries. The first is the global spin flip
\begin{equation}
    \hat{\Pi} = \prod_{i=0}^{N-1} \hat{\sigma}_i^x \,,
\end{equation}
which commutes with Eq.~\eqref{eq:H_albash} because $H_\mathrm{cost}$ only contains ZZ interactions and no local fields (whereas driver and catalyst only contain Pauli-X terms and thus trivially commute with $\hat{\Pi}$). The second symmetry is the exchange of the two rings, implemented by the
ring-swap operator $\hat{P}_\mathrm{swap}$ that exchanges sites $i$ and
$i + N/2$ for all $i = 0, \ldots, N/2 - 1$. The graph of
Fig.~\ref{fig:local_ising_sketch} and its coupling pattern are manifestly
invariant under this exchange, so
$[\hat{P}_\mathrm{swap}, H_\mathrm{cost}] = 0$. Writing a computational-basis
state as $\ket{a, b}$ with $a, b \in \{0, 1\}^{N/2}$ the bitstrings on the
upper and lower rings, respectively, the two symmetry operators act as
\begin{align}
    \hat{\Pi}\ket{a, b} &= \ket{\bar{a}, \bar{b}} \,, \\
    \hat{P}_\mathrm{swap}\ket{a, b} &= \ket{b, a} \,,
\end{align}
where $\bar{a}$ denotes the bitwise complement of $a$. Both operators square to
the identity and commute with each other, generating the Abelian group
\begin{equation}
    G = \{1,\, \hat{\Pi},\, \hat{P}_\mathrm{swap},\,
          \hat{\Pi}\hat{P}_\mathrm{swap}\} \cong \mathbb{Z}_2 \times \mathbb{Z}_2 \,.
\end{equation}
The driver Hamiltonian and the catalyst Hamiltonian in
Eq.~\eqref{eq:H_albash} share these symmetries, so $G$ is a
symmetry of the full time-dependent annealing Hamiltonian at all $s$ and
$\lambda$.

\paragraph{Reduction to the symmetric sector.}
Since $G$ commutes with the Hamiltonian throughout the annealing protocol, the Hilbert
space decomposes into invariant sectors labeled by the simultaneous eigenvalues
$(\pm 1, \pm 1)$ of $(\hat{\Pi}, \hat{P}_\mathrm{swap})$. The initial state
$\ket{+}^{\otimes N}$ is the uniform superposition over all
computational-basis states and is therefore a $+1$ eigenstate of both
operators. The dynamics consequently remains confined to the $(+1, +1)$
sector, which we refer to as the \emph{symmetric sector}, and it is sufficient
to diagonalize the Hamiltonian within this subspace.
 
An orthonormal basis of the symmetric sector is obtained by partitioning the
$2^N$ computational-basis states into orbits under the action of $G$ and
assigning to each orbit $\mathcal{O}$ the normalized equal superposition
\begin{equation}
    \ket{\mathcal{O}} = \frac{1}{\sqrt{|\mathcal{O}|}}
    \sum_{\ket{x} \in \mathcal{O}} \ket{x} \,.
\end{equation}
Each $\ket{\mathcal{O}}$ is invariant under every element of $G$, orbits with
disjoint supports yield orthogonal states, and any vector in the symmetric
sector can be expanded in this basis. 

The dimension $d$ of the symmetric
sector is therefore equal to the number of orbits, which, by Burnside's lemma,
equals the average number of computational-basis states fixed by the elements
of $G$.
A direct enumeration of fixed points gives the following. The identity fixes
all $2^N$ basis states. The spin flip $\hat{\Pi}$ fixes none, since no
bitstring equals its own complement. The ring-swap $\hat{P}_\mathrm{swap}$
fixes the $2^{N/2}$ states with $a = b$, and the combined operation
$\hat{\Pi}\hat{P}_\mathrm{swap}$ fixes the $2^{N/2}$ states with $a = \bar{b}$.
Burnside's lemma then yields
\begin{align}
\label{eq:local_ising_sector_dimension}
    d &= \frac{1}{|G|} \sum_{g \in G} |\mathrm{Fix}(g)| \nonumber\\
      &= \frac{1}{4}\bigl(2^N + 0 + 2^{N/2} + 2^{N/2}\bigr)
      = 2^{N-2} + 2^{N/2-1} \,,
\end{align}
which approaches $2^N/4$ at large $N$. The reduction is by a constant factor
only, so that the symmetric-sector dimension remains exponential in $N$.
Nonetheless, it extends the range of system sizes accessible to exact
diagonalization by about two qubits, which is still useful for the finite-size
scaling analysis of Sec.~\ref{sec:localising}.

\paragraph{Observables and MPS-based estimation of the SRE.}
We characterize the local Ising model by the same three observables considered
for the $p$-spin model, namely the minimum energy gap, the bipartite
entanglement entropy of Eq.~\eqref{eq:entanglement-entropy}, and the stabilizer
R\'enyi entropy of Eq.~\eqref{eq:SRE}. For the bipartite entanglement entropy, 
we reconstruct the full wavefunction in the $2^N$-dimensional computational
basis from its symmetric-sector representation and compute the reduced density
matrix for a balanced bipartition into the upper and lower rings. This step is
numerically inexpensive compared to the diagonalization itself and therefore
does not restrict the accessible system sizes further.

The stabilizer R\'enyi entropy, by contrast, requires in principle the
evaluation of $4^N$ Pauli-string expectation values, which becomes prohibitive
beyond small system sizes in the absence of permutation symmetry. To access
larger $N$, we represent the instantaneous ground state as a matrix product
state (MPS) and estimate the SRE via Monte Carlo sampling over the Pauli group,
following the methods of Ref.~\cite{Tarabunga2023}. In our implementation, we use bond
dimension $\chi = 16$ for $N \le 14$ and $\chi = 32$ for $N \ge 16$. The SRE
estimate is obtained by averaging over $10$ independent Monte Carlo runs, with
$10^4$ samples per run for $N < 16$ and $3 \times 10^4$ samples per run for
larger systems.

\paragraph{Discretization of the annealing parameter.}

As for the $p$-spin model (Appendix~\ref{app:p-spin}), the annealing parameter $s$ is
discretized non-uniformly, with finer resolution in the vicinity of the
relevant extrema, namely the minima of the energy gap and the maxima of
the bipartite entanglement entropy and the SRE. For the stoquastic
protocol ($\lambda = 0$), a uniform step $\Delta s = 10^{-4}$ is
sufficient for all three observables. For the non-stoquastic protocol
($\lambda = 4$), the grid is refined within the intervals that contain
the extrema. The energy gap is evaluated with $\Delta s = 5\times10^{-5}$
for $N = 8, 10$ and $\Delta s = 10^{-5}$ for $N \ge 12$. The SRE uses the
same resolutions, except at the largest sizes $N = 18, 20$, where
$\Delta s = 5\times10^{-5}$ is employed. The entanglement entropy is
evaluated with $\Delta s = 10^{-5}$ for all accessible system sizes. We
verified that these resolutions yield converged estimates of the
extremal values entering the finite-size analysis of
Sec.~\ref{subsec:local_ising_scaling}.

\section{Microscopic description of magnetization transitions in the local Ising model} 
\label{app:local-ising-effective-model}

In this appendix, we provide a microscopic interpretation of the plateau structure
observed in the global magnetization of the local Ising model
(Fig.~\ref{fig:magnetization}). We construct an effective Hamiltonian valid in
the regime of large catalyst strength and early annealing time, in which the cost
Hamiltonian contributes only perturbatively. Within this effective description,
each plateau corresponds to a classical spin configuration of an $XX$-type model. 
The boundaries between plateaus follow from explicit classical level crossings
between these configurations, which are broadened into avoided crossings by non-commuting terms appearing at subsequent orders.  

\subsection{Effective Hamiltonian in the large-$\lambda$, small-$s$ regime}
We start from the full annealing Hamiltonian, Eq.~\eqref{eq:H_albash}, and consider
the regime in which the catalyst term dominates over the cost term,
$\lambda\, s(1-s) \gg s$. Keeping the product
$\tilde{\lambda}\equiv\lambda s$ fixed and expanding in $s$ to leading order, the effective Hamiltonian becomes
\begin{equation}
\label{eq:H_effective_app}
\hat{H}_\mathrm{eff} = -\sum_i \hat{\sigma}_i^x
+ \tilde{\lambda}\sum_{\langle i,j\rangle}\hat{\sigma}_i^x\hat{\sigma}_j^x
+ \mathcal{O}(s)\,.
\end{equation}
To leading order, the cost Hamiltonian drops out and the dynamics is governed by
an Ising model in $X$ direction with competing on-site longitudinal field and nearest-neighbor couplings. Since all terms in Eq.~\eqref{eq:H_effective_app} are diagonal in the
$\hat{\sigma}^x$ eigenbasis and commute pairwise, the ground state is the
minimum-energy eigenstate over classical spin configurations $\{s_i\}$ with
$s_i=\pm 1$, whose energy reads
\begin{equation}
\label{eq:E_effective_app}
E(\{s_i\}) = -\sum_i s_i
+ \tilde{\lambda}\sum_{\langle i,j\rangle} s_i s_j\,.
\end{equation}

The competition between the two terms is controlled entirely by $\tilde{\lambda}$.
For $\tilde{\lambda}\ll 1$, the first term dominates and the ground state is fully
polarized along $+x$. For $\tilde{\lambda}\gg 1$, the second term dominates and
favors configurations with $s_i s_j = -1$ on as many bonds as possible, i.e.,
antiferromagnetic arrangements in the $\hat{\sigma}^x$ basis with vanishing (or
minimal) magnetization. At intermediate $\tilde{\lambda}$, neither
term wins globally, and a sequence of configurations becomes successively
favored as $\tilde{\lambda}$ grows. 

For fixed large catalyst strength \(\lambda\), the parameter
\(\tilde{\lambda}=\lambda s\) should therefore be interpreted as a rescaled
early-time coordinate. The regime \(\tilde{\lambda}\ll 1\) corresponds to the
very beginning of the sweep, \(s\ll 1/\lambda\), where the transverse field
dominates. The plateau transitions occur at \(s=O(1/\lambda)\), where
\(\tilde{\lambda}=O(1)\). Finally, \(\tilde{\lambda}\gg 1\) corresponds to
times later within the small-\(s\) window, \(1/\lambda\ll s\ll 1\), but not to the
end of the annealing sweep. In this regime the \(XX\) catalyst term dominates the
effective Hamiltonian, while the cost Hamiltonian remains perturbative as long
as \(s=\tilde{\lambda}/\lambda\ll 1\).

\subsection{Classical configurations and plateau states}
\begin{figure}
    \centering
    \includegraphics[width=\linewidth]{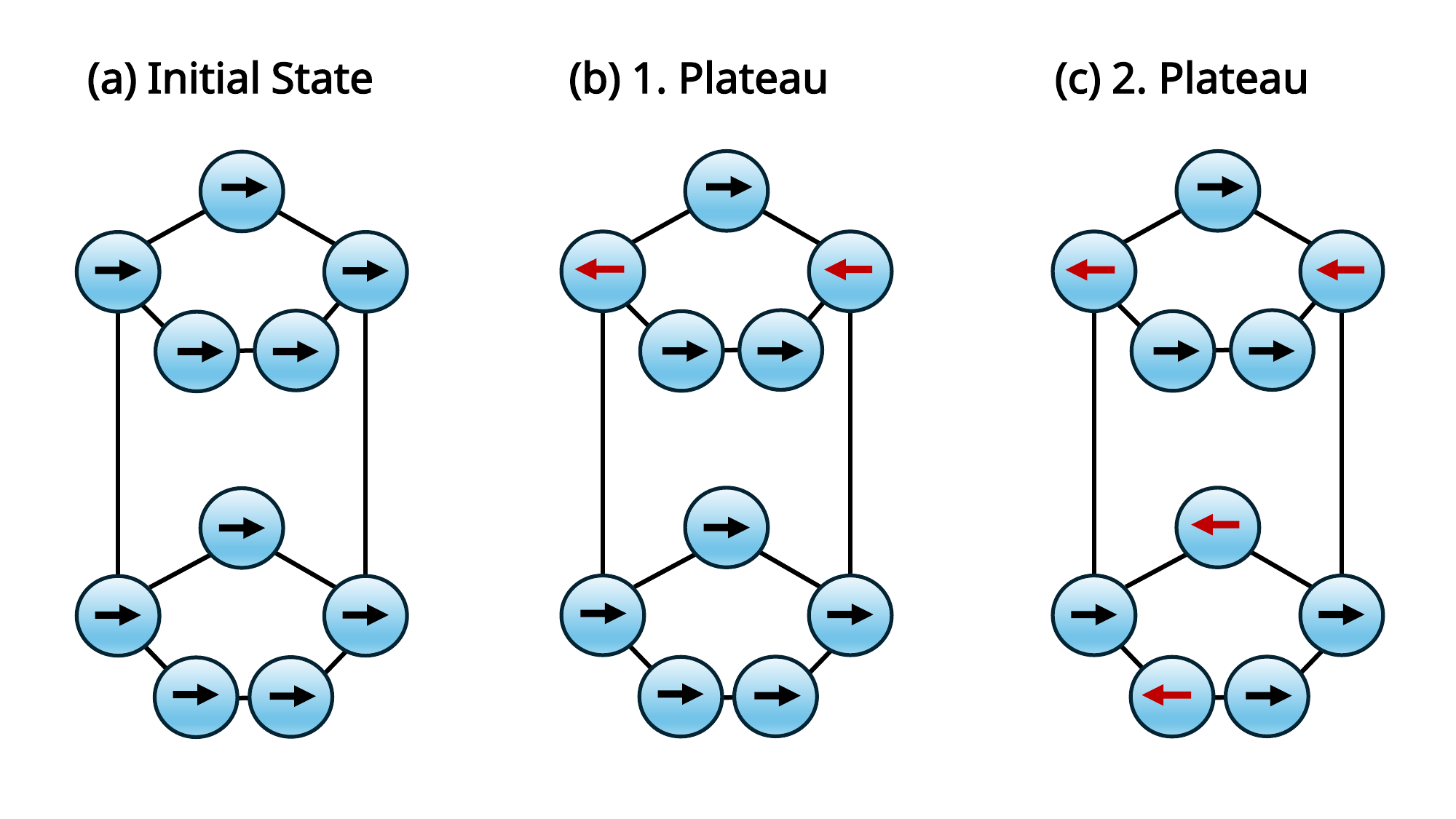}
    \caption{Representative spin configurations of the plateau states of the
    local Ising model for $N=10$, shown in the $\hat{\sigma}^x$ basis.
    (a) Fully polarized initial state, $s_i=+1$ for all $i$.
    (b) First intermediate plateau, with two non-adjacent edge spins flipped
    to $s_i=-1$.
    (c) Final plateau, reached at the end of the small-$s$ window and
    characterized by strongly reduced $X$-magnetization (full
    antiferromagnetic alignment is obstructed by frustration on the odd-length
    rings).
    Arrows indicate the classical value $s_i=\pm 1$ at each site. Only one
    representative configuration is shown per plateau. The true plateau state
    is a symmetric superposition of all configurations (sharing the same energy) related by the
   spatial symmetries of the effective Hamiltonian (Eq.~\eqref{eq:H_effective_app}).}
    \label{fig:appendix_local_ising_sketch_N=10}
\end{figure}

\begin{figure}
    \centering
    \includegraphics[width=\linewidth]{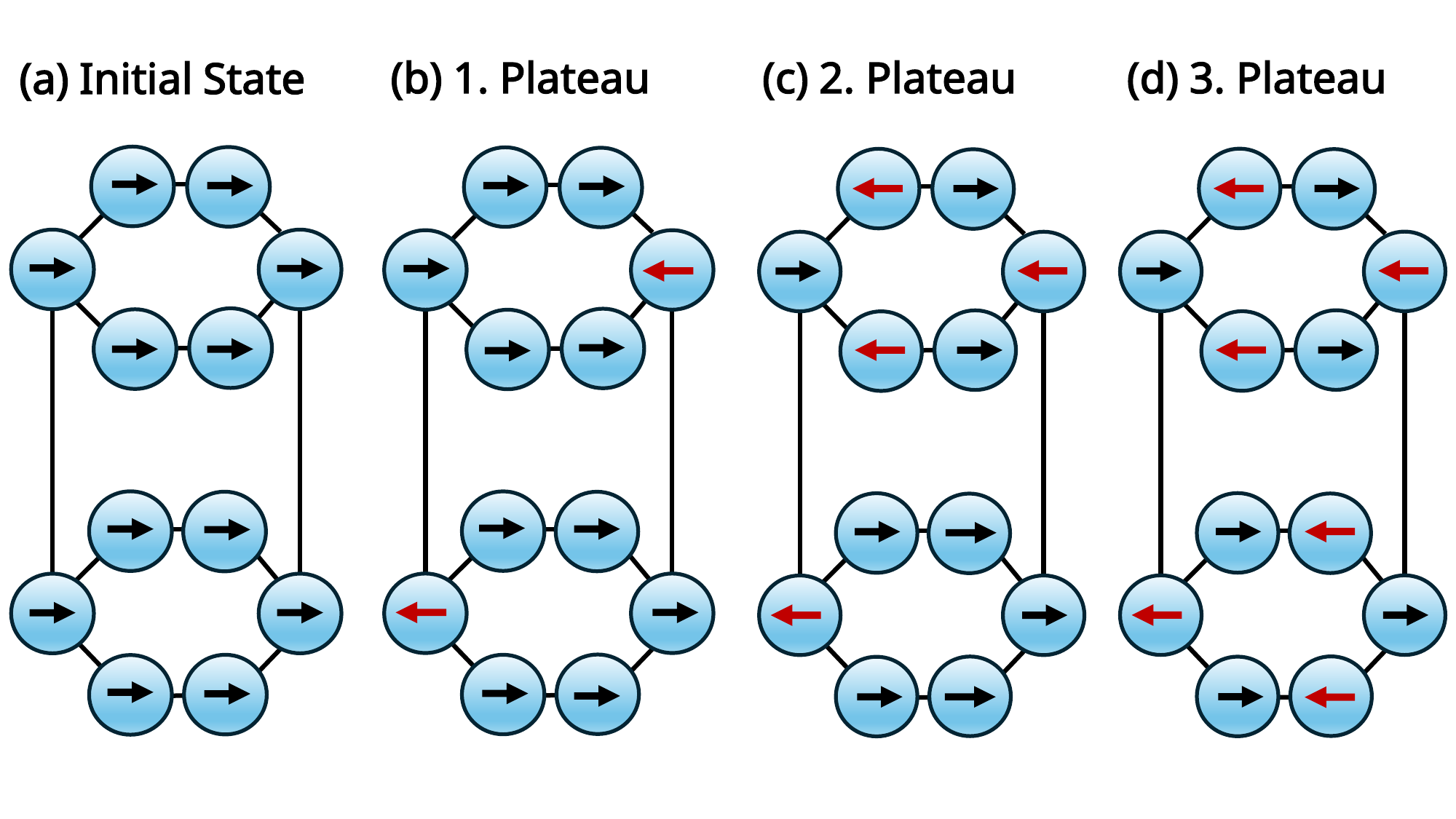}
    \caption{Same as Fig.~\ref{fig:appendix_local_ising_sketch_N=10}, but for
    $N=12$. In this case, the longer bipartite rings support an additional
    intermediate configuration between the first plateau and the final state. Moreover, the two rings have an even length, permitting for full antiferromagnetic alignment in the last plateau.}
    \label{fig:appendix_local_ising_sketch_N=12}
\end{figure}

Let $\mathcal{G}$ denote the two-ring graph defined by $H_\mathrm{cost}$
(Fig.~\ref{fig:local_ising_sketch}), with $|E(\mathcal{G})|=N+2$ edges. The graph
contains four ``edge spins'' (two per ring, at diametrically opposite
positions) of coordination number three, and $N-4$ ``bulk spins'' of
coordination number two. Starting from the fully polarized configuration
$|{+}\rangle^{\otimes N}$ with all $s_i = +1$, a different discrete family of configurations
becomes energetically favored as $\tilde{\lambda}$ grows, each obtained by
flipping a specific subset of spins. These configurations correspond to the
plateaus observed in Fig.~\ref{fig:magnetization} and are sketched in
Figs.~\ref{fig:appendix_local_ising_sketch_N=10}
and~\ref{fig:appendix_local_ising_sketch_N=12} for $N=10$ and $N=12$, respectively.

For a configuration obtained from the fully polarized state by flipping a
subset \(S\) of spins, the field contribution is changed only by the number
of flipped spins. Since each flipped spin changes \(s_i=+1\) to \(s_i=-1\),
the first term in Eq.~\eqref{eq:E_effective_app} changes from \(-N\) to
\(-N+2|S|\). The interaction term is modified only on bonds connecting a
flipped spin to an unflipped spin: on such a bond, \(s_i s_j\) changes from
\(+1\) to \(-1\), lowering the \(XX\)-bond contribution by
\(2\tilde{\lambda}\). Bonds whose endpoints are both flipped, or both
unflipped, remain unchanged. If \(B(S)\) denotes the number of bonds with
one endpoint in \(S\) and the other outside \(S\), the energy then reads
\begin{equation}
\label{eq:E_flip_general_app}
E(S) = -\bigl[N - 2|S|\bigr]
+ \tilde{\lambda}\bigl[(N+2) - 2B(S)\bigr] .
\end{equation}
Thus, energetically favorable flipped-spin configurations are those that
gain many antiferromagnetically aligned \(XX\) bonds while paying the
energy cost associated with the flipped spins.
We can separate this general formula into the following special cases. 
We note that within the +1 parity sector that contains the initial state, only configurations with an even number of flipped spins are accessible, so the plateau sequence proceeds in steps of two flipped spins (compare with the discussion of symmetries in Appendix~\ref{app:local-ising}).

\paragraph{Fully polarized state.}
All spins are aligned with the $X$ magnetic field, $S=\emptyset$. Every bond
contributes $+\tilde{\lambda}$, yielding
\begin{equation}
\label{eq:E_pol_app}
E_\mathrm{pol} = -N + \tilde{\lambda}(N+2)\,.
\end{equation}
This state is favored at small $\tilde{\lambda}$, i.e., at the very beginning of the annealing sweep. 
 
\paragraph{Final antiferromagnetic state.}
For bipartite rings ($N/2$ even), all bonds can be simultaneously satisfied with
$s_i s_j = -1$ and the total magnetization vanishes, giving
\begin{equation}
\label{eq:E_AFM_app}
E_\mathrm{AFM} = -\tilde{\lambda}(N+2)\,.
\end{equation}
For odd rings ($N/2$ odd, e.g., $N=10$), full antiferromagnetic alignment is
obstructed by frustration, so the final-state energy is slightly higher
than Eq.~\eqref{eq:E_AFM_app}.
This configuration is favored once $\tilde{\lambda}=\lambda s\gg 1$, but still $s\ll 1$, i.e., in an intermediate regime of an annealing sweep at $\lambda\gg 1$ fixed. 

\paragraph{First intermediate plateau.}
Because edge spins have higher coordination, each contributes more strongly to
the $XX$ term when flipped, and is therefore favored to flip first as
$\tilde{\lambda}$ increases. The configuration corresponding to the first
plateau is obtained by flipping two edge spins, chosen such that they are not
directly connected (see
Figs.~\ref{fig:appendix_local_ising_sketch_N=10}(b)
and~\ref{fig:appendix_local_ising_sketch_N=12}(b)). For this choice, 
$|S|=2$ and $|\partial S|=6$ (the $2\times 3 = 6$ bonds incident to the two
flipped edge spins, none of which lies between them), giving
\begin{equation}
\label{eq:E_plateau1_app}
E_\mathrm{plateau,1} = -N + 4 + \tilde{\lambda}(N-10)\,.
\end{equation}
Equation~\eqref{eq:E_plateau1_app} is independent of which non-adjacent pair of
edge spins is chosen, since the graph is symmetric under parity
$\hat{\Pi}=\prod_i \hat{\sigma}_i^x$ and ring-swap $\hat{P}_\mathrm{swap}$. The
actual plateau state is a symmetric superposition over all representatives
related by these operations, each sharing the same classical energy. This
superposition is responsible for the depolarization of the four connection
spins visible in the spin-resolved magnetization profile of
Fig.~\ref{fig:magnetization}: the individual representatives have fully
polarized edge spins with $s_i = \pm 1$, but their symmetric superposition has
vanishing expectation value $\langle \hat{\sigma}_i^x \rangle = 0$ at each
connection site, while the bulk spins remain polarized in every representative
and therefore in the superposition.

\paragraph{Further plateaus ($N>10$).}
For $N=8$ and $N=10$, Eq.~\eqref{eq:E_plateau1_app} describes the unique
intermediate configuration between the polarized and the final states. For
$N>10$, additional plateaus appear, obtained by progressively flipping further
pairs of spins while keeping $|\partial S|$ as large as possible. As a
representative example, for $N=12$ the second intermediate plateau
(Fig.~\ref{fig:appendix_local_ising_sketch_N=12}(c)) has $|S|=4$ flipped spins
and $|\partial S|=10$ cut bonds, giving from
Eq.~\eqref{eq:E_flip_general_app}
\begin{equation}
\label{eq:E_plateau2_app}
E_\mathrm{plateau,2}^{(N=12)} = -4 - 6\tilde{\lambda}\,.
\end{equation}

\subsection{Boundaries between plateaus}
From the level crossings between the classical energies above, we can identify the boundaries between plateaus that are visible in the $(\lambda, s)$ plane at large $\lambda$, see Fig.~\ref{fig:magnetization}. Equating $E_\mathrm{pol}$ and
$E_\mathrm{plateau,1}$ yields, independently of $N$,
\begin{equation}
\tilde{\lambda}_1 = \frac{1}{3}
\qquad\Longleftrightarrow\qquad
s_1 = \frac{1}{3\lambda}\,,
\end{equation}
marking the transition out of the fully polarized configuration. 

Equating $E_\mathrm{plateau,1}$ with
$E_\mathrm{AFM}$ (Eq.~\eqref{eq:E_AFM_app}) further yields
\begin{equation}
\tilde{\lambda}_2 = \frac{1}{2}
\qquad\Longleftrightarrow\qquad
s_2 = \frac{1}{2\lambda}\,,
\end{equation}
which for $N\in\{8,10\}$ identifies the transition time to the final configuration
directly. For $N=12$, inserting Eq.~\eqref{eq:E_plateau2_app} one finds that
both $E_\mathrm{plateau,1}=E_\mathrm{plateau,2}$ and
$E_\mathrm{plateau,2}=E_\mathrm{AFM}$ are solved at the same value
$\tilde{\lambda}=1/2$, so that within this strictly classical analysis the
second plateau is degenerate with the first plateau and the final state at a
single point and is never the unique minimum.

\begin{figure}
    \centering
    \includegraphics{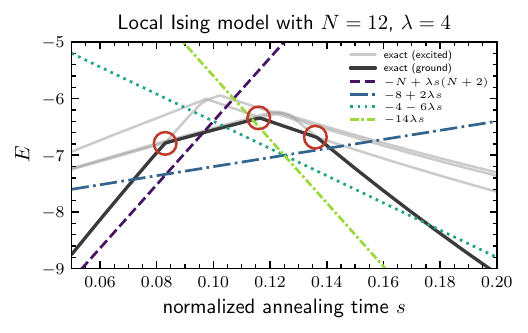}
    \caption{Low-lying energy spectrum of the local Ising model with $N=12$
    and $\lambda=4$ as a function of the normalized annealing time
    $s\in[0.05,0.2]$. Grey curves show the numerically exact spectrum (ground
    state in dark grey), while the colored lines show the analytical plateau
    energies of Eqs.~\eqref{eq:E_pol_app}, \eqref{eq:E_plateau1_app},
    \eqref{eq:E_plateau2_app} and~\eqref{eq:E_AFM_app}. Red circles mark the
    three avoided crossings at $s\approx 0.083,\,0.116,\,0.136$. 
    Terms subleading in $s$ shift the exact energies upwards and split the single exact crossing predicted from the classical effective Hamiltonian at $s=0.125$ into two avoided crossings at $s\approx 0.116,\,0.136$.  
    }
    \label{fig:spectrum_local_ising}
\end{figure}

\subsection{Comparison with the exact spectrum}
\label{subsec:app_spectrum_comparison}
 
Figure~\ref{fig:spectrum_local_ising} compares the analytical predictions above
with the numerically exact low-lying spectrum for $N=12$ and $\lambda=4$, in
the window $s\in[0.05,0.2]$ where all relevant rearrangements take place. The
colored lines show the plateau energies, Eqs.~\eqref{eq:E_pol_app},
\eqref{eq:E_plateau1_app}, \eqref{eq:E_plateau2_app}, 
and~\eqref{eq:E_AFM_app}, while the dark and light grey curves show the exact
ground- and excited-state energies, respectively. Three avoided crossings,
marked by red circles at $s\approx 0.083,\,0.116,\,0.136$, separate four
windows of the spectrum whose ground-state slopes agree with those of the
analytical predictions, in the order prescribed by the analysis of the
previous section. The first avoided crossing is centered close to the
analytical value $s_1 = 1/(3\lambda)\approx 0.083$, while the remaining two
are to the left and right of the degenerate classical value $s_2 = 1/(2\lambda)= 0.125$.

Two systematic deviations from the classical analysis are worth noting. First,
the exact ground-state energy lies above the analytical predictions by an
offset that depends weakly on $s$. This offset reflects contributions of order
$s$ neglected in Eq.~\eqref{eq:H_effective_app}: at $\lambda=4$ and
$s\sim 0.1$ the cost Hamiltonian is not entirely negligible, and it shifts the energies upwards without altering their
slopes in the observed window. Second, the second plateau state
is degenerate rather than uniquely favored in the classical description at
$\tilde{\lambda}=1/2$, yet emerges as the exact ground state over the finite
window between the second and third avoided crossings. The lifting of the
classical degeneracy is again a consequence of the subleading cost
Hamiltonian, which selects the second-plateau configuration in this window.
Apart from these two effects, the analytical plateau structure accurately
captures the location and ordering of the transitions (the analytical prediction becomes more accurate in the large $\lambda$ limit). This demonstrates that
the phase diagram observed in Fig.~\ref{fig:magnetization} in the
large-$\lambda$, small-$s$ regime admits a quantitative microscopic
interpretation in terms of competing classical configurations of the effective
Hamiltonian, Eq.~\eqref{eq:H_effective_app}, with the full Hamiltonian
providing systematic corrections that decide the outcome at the classical
degeneracy point.

\section{Quantum resource barriers in the local Ising model} 
\label{app:local_ising_barriers}

This appendix complements the finite-size scaling analysis of Sec.~\ref{subsec:local_ising_scaling} by examining the structure of the entanglement and SRE barrier profiles along vertical annealing sweeps (i.e., at fixed strength $\lambda$ of the non-stoquastic term) of the geometrically local Ising model. The goal is to illustrate the features that motivate the restrictions adopted in Sec.~\ref{subsec:local_ising_scaling}, namely the focus on $\lambda = 4$ and the use of the global maximum as the only clear extremum.

\begin{figure*}
    \centering
    \includegraphics[width=1\linewidth]{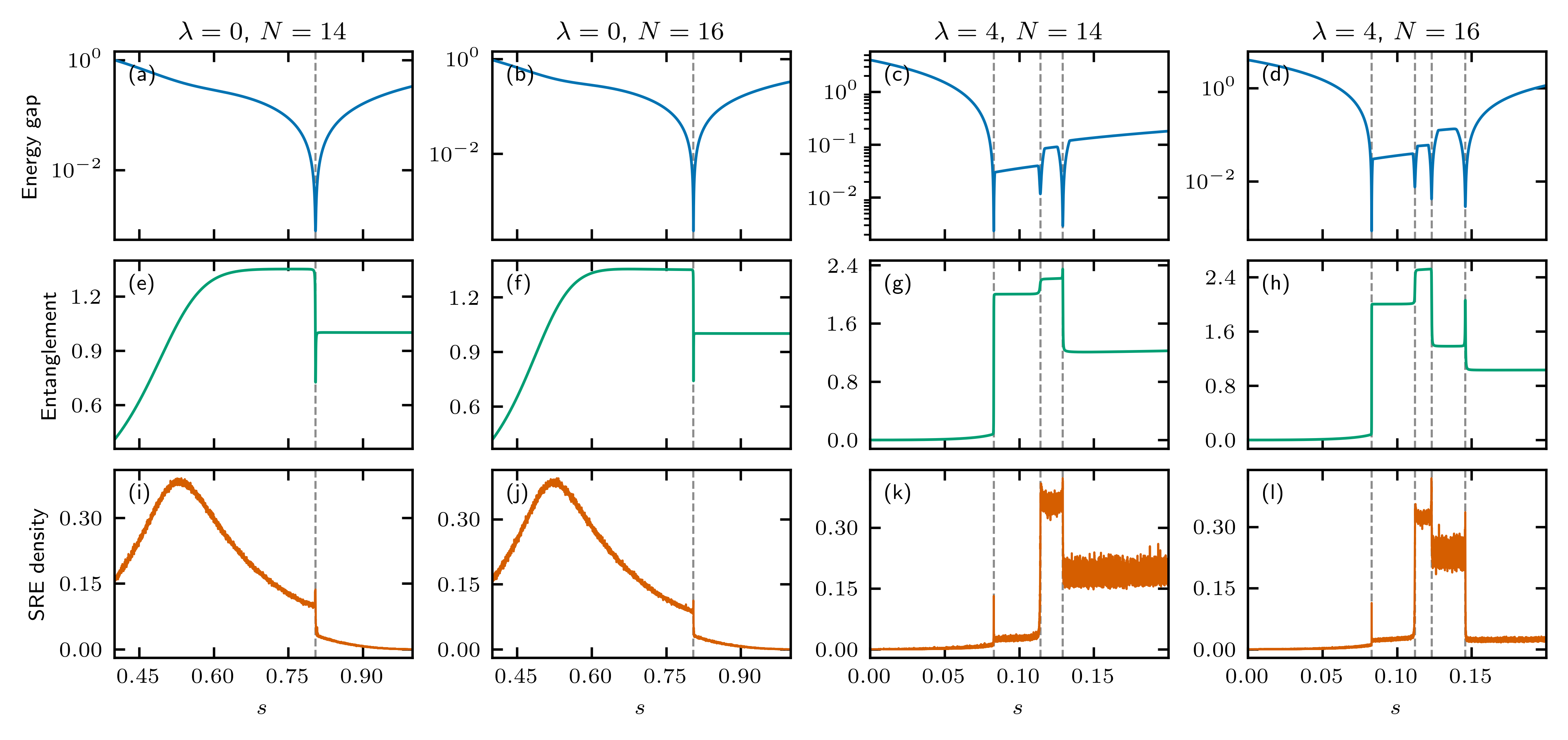}
    \caption{Energy gap (top row), bipartite entanglement entropy (middle row), and stabilizer R\'enyi entropy density (bottom row) of the instantaneous ground state along annealing sweeps for the geometrically local Ising model (see Sec.~\ref{sec:localising}), for the stoquastic ($\lambda = 0$, columns 1 and 2) and non-stoquastic ($\lambda = 4$, columns 3 and 4) protocols, and for system sizes $N = 14$ (columns 1 and 3) and $N = 16$ (columns 2 and 4). Vertical gray dashed lines mark the locations of the local minima of the energy gap and are reproduced at the same values of $s$ on the entanglement and SRE density panels for comparison. At $\lambda = 0$, a single gap minimum is present, accompanied at approximately the same location by a sharp local minimum of the entanglement entropy and a sharp local maximum of the SRE density. 
    At $\lambda = 4$, three local gap minima are visible at $N = 14$ and four at $N = 16$; at each of them, the entanglement and SRE density show transitions between plateau values, in some cases accompanied by sharp local maxima. The statistical fluctuations visible on the SRE density plateaus at $\lambda = 4$ originate from the MPS Monte Carlo estimation described in Appendix~\ref{app:local-ising}.}
    \label{fig:local-ising-barriers}
\end{figure*}

Figure~\ref{fig:local-ising-barriers} shows the energy gap, the bipartite entanglement entropy, and the SRE density along the annealing sweep for the stoquastic protocol ($\lambda = 0$, two leftmost columns) and the non-stoquastic protocol ($\lambda = 4$, two rightmost columns), for system sizes $N = 14,16$. The vertical gray dashed lines mark the locations of the local minima of the energy gap and are reproduced at the same values of $s$ on the entanglement and SRE density panels to ease visual comparison across observables.

For the stoquastic protocol ($\lambda = 0$), the energy gap displays a single minimum. At approximately the same annealing time $s$, the entanglement entropy develops a pronounced local minimum, intervening between a sharp drop from one plateau value to another one.  
The SRE density exhibits a similar sharp feature with a local maximum at the same $s$ value. Notably, this local maximum in the SRE density is not the global maximum. At an earlier annealing time $s\approx 0.55$, the SRE density already exhibits a much larger and broader maximum. Interestingly, this feature has no analogue in either energy gap or entanglement. The observed structure matches the discussion of the parameter-space picture of Fig.~\ref{fig:local-ising-grids}: in the stoquastic regime, the annealing dynamics is dominated by a single rapid rearrangement of the ground state, producing one unambiguous sharp local extremum in each observable. The qualitative structure is identical at $N = 14$ and $N = 16$.

For the non-stoquastic protocol ($\lambda = 4$), the energy gap instead develops multiple local minima within a narrow window of small $s$, consistent with the sequence of avoided crossings between competing plateau states explained in Appendix~\ref{app:local-ising-effective-model}. Three local minima are visible at $N = 14$, and four at $N = 16$. At each local gap minimum, the entanglement entropy and the SRE density display a clear feature, but its nature varies between a local maximum and a sharp transition between two plateau values without any local extremum, depending on the specific avoided crossing. 
This behavior complicates the identification of individual local extrema beyond the global maximum for a clear scaling analysis. First, the number of features depends on $N$. Second, often the location of an avoided crossing lies at a sharp transition between two plateaus without any isolated peak, hindering the unambiguous identification of a value associated with the gap closing. Third, the Monte Carlo estimate of the SRE (Appendix~\ref{app:local-ising}) introduces statistical fluctuations visible on top of the plateaus. For these reasons, in the finite-size analysis of Sec.~\ref{subsec:local_ising_scaling} we restrict our attention to the scaling of the global maximum of entanglement and SRE attained along the annealing path, which remains well-defined for all accessible $N$. By contrast, in the $p$-spin model the sinusoidal schedules of Sec.~\ref{subsec:p-spin_barriers} produce a single dominant peak in each quantum resource (Fig.~\ref{fig:p-spin-2}), allowing for a cleaner extraction of the barrier heights and a more direct finite-size scaling analysis.

\end{document}